\documentclass[11pt]{article}
\usepackage[latin9]{inputenc}
\usepackage{amsfonts}
\usepackage{amsmath}
\usepackage{amssymb}
\usepackage{indentfirst}
\usepackage{graphicx}
\usepackage[colorlinks]{hyperref}
\usepackage{cite}

\numberwithin{equation}{section}
\begin{document}

\begin{titlepage}
\vspace{3cm}
\baselineskip=24pt

\begin{center}
\textbf{\LARGE{On the supersymmetry invariance of flat supergravity with boundary}}
\par\end{center}{\LARGE \par}

\begin{center}
	\vspace{1cm}
	\textbf{Patrick Concha}$^{\ast}$,
	\textbf{Lucrezia Ravera}$^{\ddag}$,
	\textbf{Evelyn Rodríguez}$^{\dag}$
	\small
	\\[5mm]
	$^{\ast}$\textit{Instituto
		de Física, Pontificia Universidad Católica de Valparaíso, }\\
	\textit{ Casilla 4059, Valparaiso-Chile.}
	\\[2mm]
                 $^{\ddag}$\textit{INFN, Sezione di Milano, }\\
	\textit{ Via Celoria 16, I-20133 Milano-Italy.}
	\\[2mm]
	$^{\dag}$\textit{Departamento de Ciencias, Facultad de Artes Liberales,} \\
	\textit{Universidad Adolfo Ibáñez, Viña del Mar-Chile.} \\[5mm]
	\footnotesize
	\texttt{patrick.concha@pucv.cl},
	\texttt{lucrezia.ravera@mi.infn.it},
	\texttt{evelyn.rodriguez@edu.uai.cl}
	\par\end{center}
\vskip 20pt
\begin{abstract}
\noindent
The supersymmetry invariance of flat supergravity {(i.e., supergravity in the absence of any internal scale in the Lagrangian)} in four dimensions on a manifold with non-trivial boundary is explored. Using a geometric approach we find that the supersymmetry invariance of the Lagrangian requires to add appropriate boundary terms. This is {achieved} by considering additional gauge fields to the boundary without modifying the bulk Lagrangian. We also construct an enlarged {supergravity model from which}, in the vanishing cosmological constant limit, flat supergravity with {a non-trivial} boundary emerges properly.

\end{abstract}
\end{titlepage}\newpage {}

\section{Introduction}

The presence of a boundary in (super)gravity theories has been of particular
interest {in the} last 40 years \cite{York, GH, BY, HW}. The addition of
boundary terms plays an important role in the so-called AdS/CFT {duality \cite{Maldacena, GKP, Witten, AGMOO, HF, CM, AC}}. Such duality between a quantum field
theory living on the boundary and a string theory on asymptotically AdS
spacetime implies in the supergravity limit a one to one correspondence
between fields of the bulk supergravity theory and quantum operators in the
boundary CFT. This requires to consider proper boundary conditions for the
supergravity fields which act as sources for the operators of the CFT.
Interestingly, the divergences of the bulk metric can be cancelled by adding
appropriate counterterms at the boundary (holographic renormalization) \cite{BK, BVV, VV, Boer, HSS, Skenderis, LS, HP}. More recent results of proper counterterms in gravity theories beyond Einstein theory have been developed in \cite{ABD, BRadu, AACM, ABCR}.

The inclusion of boundary terms in supergravity has been studied by diverse
authors {in \cite{EKK, Moss, NV, B, NRVW, BN, BN1, GN, HPSS, AD, PKS, CIRR, FPPW, BCMS, AAGT, ACDT, BR}.} In particular, in \cite{NV, B, BN1, GN} it was pointed out
that, unlike the Gibbons-Hawking prescription \cite{GH}, the supersymmetry
invariance of a supergravity action should be satisfied without imposing
Dirichlet boundary conditions. Interestingly, for the $\mathcal{N}=1$
three-dimensional supergravity, it was proven that the boundary term
reproduces not only the Gibbons-Hawking-York boundary term but also the
counterterm allowing to regularize the action \cite{GN}.

Recently, the authors of \cite{AD} have shown that the supersymmetry
invariance of $\mathcal{N}=1$ and $\mathcal{N}=2$ four-dimensional
supergravity with cosmological constant requires the presence of topological
terms. Such result can be seen as the supersymmetric extension of those
obtained in AdS gravity in which the addition of the topological
Gauss-Bonnet term allows to regularize the action \cite{ACOTZ, ACOTZ2, MOTZ,
Olea, JKMO}. Subsequently{,} in \cite{CIRR, BR}, using the same geometric
approach used in \cite{AD}, it was shown that the supersymmetric extension
of a Gauss-Bonnet like gravity is required to restore the supersymmetry
invariance of enlarged supergravities in the presence of a non-trivial
boundary.\footnote{{In presence of a non-trivial boundary of spacetime, that is when the boundary is not at infinity, the fields do not asymptotically vanish, and this has some consequences on the invariances of the theory, in particular on supersymmetry invariance.}} Interestingly, the full supergravity actions obtained in \cite{AD, CIRR, BR} can be rewritten in terms of the super curvatures à la MacDowell{-}Mansouri \cite{MM}.

{Boundary conditions imposed at a finite value of the radial coordinate have been studied in $D=4$ supergravity with zero cosmological constant in \cite{NV, NRVW, BN}. However, the limit case of vanishing cosmological constant in the presence of a non-trivial boundary remains poorly explored.}
In particular, the flat supergravity Lagrangian is not supersymmetric when a {non-trivial} boundary is considered {and, on} the other hand, a flat limit cannot be naively applied to the MacDowell{-}Mansouri Lagrangian since it reduces only to boundary terms:%
\begin{equation}
\mathcal{L}=\epsilon _{abcd}\mathcal{R}^{ab}\mathcal{R}^{cd}+D\bar{\psi}%
\gamma _{5}D\psi \,.
\end{equation}%
This can be directly seen from the super curvatures $\mathcal{R}^{ab}$ and $%
D\psi $ of the Poincaré superalgebra which do not allow the presence of the
Einstein-Hilbert (EH) neither the Rarita Schwinger (RS) terms in the
Lagrangian. This inconvenient appears only in presence of a boundary since
the bulk flat Lagrangian can be recovered directly from the bulk $OSp\left(
4|1\right) $ supergravity as a vanishing cosmological constant limit.

{Let us specify that, here as well as in the sequel of this work, with ``flat supergravity'' we mean supergravity in the absence of any internal scale in the Lagrangian.}

In this paper, we restore the supersymmetry invariance of the
four-dimensional flat supergravity in presence of a non-trivial boundary. To
this purpose, we introduce appropriate boundary terms to the Lagrangian such
that the supersymmetry is recovered. This is
achieved by adding new bosonic and fermionic gauge fields to the boundary in
addition to the usual spin-connection, vielbein and gravitino. Interestingly, we find that the boundary values of the super curvatures are fixed by the field equations of the full Lagrangian {(understood as bulk plus boundary contributions)}. In particular, the full Lagrangian
obtained can be rewritten in {terms} of super curvatures of a particular
superalgebra known as Maxwell superalgebra \cite{BGKL}. We also present a
proper vanishing cosmological constant limit from an enlarged supergravity theory {with a non-trivial} boundary. Such enlarged supergravity Lagrangian can be rewritten à la MacDowell{-}Mansouri for a deformation of the $\mathfrak{osp}\left(4|1\right) $ superalgebra which corresponds to a new supersymmetric
extension of the AdS-Lorentz algebra \cite{SS, GKL}.

The paper is organized as follows: In Section \ref{Section2}, we present a brief review
of the geometric approach we will adopt for the formulation of supergravity in superspace. The supersymmetry invariance of a supergravity Lagrangian {in the presence of a non-trivial boundary is explicitly discussed within this framework.} Section \ref{Section3} and \ref{Section4} contain our main results.
{In particular, in} Section \ref{Section3} we explore the supersymmetry invariance of flat supergravity
in presence of a non-trivial boundary{, while in Section \ref{Section4}} we construct an
enlarged supergravity Lagrangian with a generalized cosmological constant
term using the geometric approach. We show that the flat limit can be
appropriately applied in presence of a boundary. We conclude our work with
some comments and possible future developments.

\section{Geometric approach and supersymmetry invariance in presence of a boundary}\label{Section2}

An interesting and powerful approach for constructing supergravity theories
is the geometric or rheonomic approach \cite{CDF}. The principal demand
of any supergravity theory is the invariance of the action under
supersymmetry transformations. In the rheonomic approach, a supergravity
theory is given in terms of {1-form superfields} defined on superspace {(whose basis is given by the supervielbein)}, and
the supersymmetry transformations on spacetime correspond to diffeomorphisms
in the fermionic directions of superspace. Thus, the principle of rheonomy
makes the extension from spacetime to superspace uniquely defined and
consequently allows for a geometric interpretation of the supersymmetry
rules.

In this framework, the supersymmetry invariance of the Lagrangian is
expressed by {the vanishing of the Lie derivative of the Lagrangian for
infinitesimal diffeomorphisms in the fermionic directions, that is to say}
\begin{equation}
\delta _{\epsilon }\mathcal{L}=l_{\epsilon }\mathcal{L}=\imath _{\epsilon }d
\mathcal{L}+d(\imath _{\epsilon }\mathcal{L})=0\,. \label{RC}
\end{equation}
From condition \ref{RC} it is direct to see that when a supergravity Lagrangian is considered on spacetimes without boundary, it implies that $\imath _{\epsilon }\mathcal{L}|_{\partial \mathcal{M}}=0$. Then, in the absence of a (non-trivial) boundary, $\imath _{\epsilon }d \mathcal{L} =0$ results to be a sufficient condition for supersymmetry invariance.
Nevertheless, when the background spacetime has a non-trivial boundary, the condition $%
\imath _{\epsilon }\mathcal{L}|_{\partial \mathcal{M}}=0$ becomes non-trivial. Thus, in order to verify the invariance of the Lagrangian in
presence of a non-trivial boundary, it is necessary to check it explicitly (see \cite{AD} for further details).

As it is well explained in \cite{CDF}, in the {geometric} approach the $1$-forms
fields $\mu ^{A}$ are extended from spacetime to superspace, such that the
mapping $\mu ^{A}(x)\rightarrow \mu ^{A}(x,\theta )$ is defined {by rheonomy, which amounts to the following: The superspace equations of motion are given in terms of the superspace curvatures, and can be analyzed by expanding the curvatures along the basis of $2$-forms in superspace,
\begin{equation}
R^{A}=R_{BC}^{A}\mu ^{B}\mu ^{C}\,;
\end{equation}
expanding the curvature $2$-forms $R^{A}$ in superspace along the supervielbein basis {$\lbrace V^{a},\psi \rbrace$},
\begin{equation}
R^{A}=R_{bc}^{A}V^{b}V^{c}+R_{a\alpha }^{A}V^{a}\psi ^{\alpha }+R_{\alpha
\beta }^{A}\psi ^{\alpha }\psi ^{\beta }\,,
\end{equation}
from the analysis of the equations of motion one finds that the outer components $R_{a\alpha }^{A}$, $R_{\alpha \beta }^{A}$ of the curvature $2$-forms in superspace are expressed, on-shell, as linear tensor combinations of the inner components $%
R_{ab}^{A}$.}\footnote{{The ``outer'' components are those having at least one index along the $\psi$ direction of superspace, while when the only non-vanishing components are along the bosonic vielbein they are called ``inner''.}}
{From the physical point of view, this means that no new
degree of freedom is introduced in the theory other than those already present on spacetime. Thus, rheonomy avoids the introduction of spurious degrees of freedom which would appear in the theory if the outer components of the curvatures were independent fields.}
Furthermore, if we also assume
Lorentz gauge invariance of the rheonomic parametrization for the curvature
2-forms and homogeneous scaling of all the terms involved, then the form of
the superspace curvatures is completely determined, except for some constant coefficients which are fixed by the Bianchi identities.

{Then, the rheonomic parametrization of the super curvatures can be equivalently read as the on-shell prescription for the contractions of the super field-strengths. On the other hand, one can prove that, in the geometric formalism, the condition for a theory (in particular, for its Lagrangian) to be supersymmetry invariant ($\imath_\epsilon d \mathcal{L} =0$) is fulfilled if one requires constraints on the
components of the curvatures (in particular, on the contractions of the super field-strengths). These requirements turn out to be the same of the on-shell prescription for the contractions of the super field-strengths arising from the rheonomic parametrization of the super curvatures. Thus, one retrieves exactly the same constraints on the curvatures as those found from the equations of motion.
Moreover, these constraints provide the supersymmetry transformation laws of
the fields on spacetime, under which the spacetime Lagrangian is invariant up to boundary
terms.}
{One can then show (see \cite{CDF} for details) that this restricted form of the curvatures in superspace implies that the supersymmetry transformations leaving the Lagrangian invariant close, in general, only on-shell (that is, transformations close only on the equations of motion).}
{Finally, the rheonomy principle is completely equivalent to the requirement of spacetime
supersymmetry (see \cite{CDF}), the latter being interpreted from the geometrical point of view as the requirement that the superspace equations of motion imply that the outer components of the super curvatures are linearly expressible in terms of the inner components.}

{Along this work, we will apply the geometric approach to restore the supersymmetry of flat supergravity in the presence of a boundary: In this context, one can introduce in a geometric way appropriate boundary terms to the Lagrangian in such a way that the action, including the boundary contributions, results to be invariant under supersymmetry transformations. In particular, since the geometric approach does not require an off-shell formulation of bulk supergravity, we shall use the on-shell formulation of supergravity (that is, transformations close only on the equations of motion).}

\section{Flat supergravity in presence of a non-trivial boundary}\label{Section3}

{The Lorentz-covariant super field-strengths in four dimensions are
\begin{align}
\mathcal{R}^{ab} & \equiv  d\omega ^{ab}+\omega _{\;c}^{a}\omega ^{cb} \,, \\
R^a & \equiv  D V^a - \frac{1}{2} \bar{\psi} \gamma^a \psi = d V^a + \omega^a_{\;b} V^b - \frac{1}{2} \bar{\psi} \gamma^a \psi \,, \label{supt} \\
\rho & \equiv  D \psi = d \psi + \frac{1}{4} \omega^{ab} \gamma_{ab} \psi \,,
\end{align}
being $\mathcal{R}^{ab}$, $R^a$, and $\rho$ the Lorentz super curvature, the supertorsion, and the gravitino super field-strength 2-forms, respectively ($D=d+\omega$ is the Lorentz covariant derivative), and it is well known that the four-dimensional flat supergravity Lagrangian%
\begin{equation}\label{FirstLagr}
\mathcal{L}_{\text{bulk}}=\epsilon _{abcd}\mathcal{R}^{ab}V^{c}V^{d}+4\bar{%
\psi}V^{a}\gamma _{a}\gamma _{5}\rho
\end{equation}%
is simply given by the EH and RS terms, without involving a cosmological
constant term.\footnote{{Notice that \eqref{FirstLagr} is written as a first-order Lagrangian, and the field equation for the spin-connection $\omega^{ab}$ implies (up to boundary terms, which will be considered subsequently in this work) the vanishing, on-shell, of the supertorsion $R^a$ defined in eq. \eqref{supt}.}} The field equations read (up to boundary terms)
\begin{align}
& \epsilon_{abcd} R^c V^d = 0 , \\
& 2 \epsilon_{abcd} \mathcal{R}^{ab} V^c + 4 \bar{\psi} \gamma_d \gamma_5 \rho = 0 , \\
& 8 V^a \gamma_a \gamma_5 \rho + 4 \gamma_a \gamma_5 \psi R^a = 0 .
\end{align}
Here, $a,b,\dots =0,1,2,3$
are Lorentz indices and $\epsilon _{abcd}$ is the four-dimensional
Levi-Civita tensor.} Note that the supergravity Lagrangian scales with $%
\mathcal{\omega }^{2}$, being $\mathcal{\omega }^{2}$ the scale-weight of
the EH term. In fact, $[\omega ^{ab}]=\omega^{0}$, $[V^{a}]=\omega^{1}$ and $[\psi]=\omega^{1/2}$.

{The Lagrangian \eqref{FirstLagr} is invariant in the bulk under supersymmetry of the super-Poincar\'{e} group.}
Then, in the {rheonomic} approach, we have that $\imath _{\epsilon }\left( d%
\mathcal{L}_{\text{bulk}}\right) =0\,$\ is satisfied. Nevertheless, the
supersymmetry invariance of the Lagrangian is not guaranteed when a boundary
is present. In particular, in presence of a {non-trivial} boundary, the
condition%
\begin{equation}
{\imath _{\epsilon }\mathcal{L}|_{\partial \mathcal{M}}=0}
\end{equation}%
is not necessarily satisfied and {requires} to be revised in order to restore
supersymmetry invariance of the theory. To this purpose, we have to modify
the bulk Lagrangian by adding terms that do not modify the dynamics but
affect only the boundary. Thus, we have to consider boundary topological
terms.

A good candidate for being considered as a boundary contribution should first
scale homogeneously. In particular, each term must have the same
scale-weight as the EH term. However, the only boundary terms that can be
constructed using the spin-connection $\omega ^{ab}$, the vielbein $V^{a}$
and the gravitino $\psi $ are%
\begin{eqnarray}
d\left( \omega ^{ab}\mathcal{R}^{cd}+\omega _{\text{ }f}^{a}\omega
^{fb}\omega ^{cd}\right) \epsilon _{abcd} &=&\mathcal{R}^{ab}\mathcal{R}%
^{cd}\epsilon _{abcd}\,, \\
d\left( \bar{\psi}\gamma ^{5}\rho \right) &=&\bar{\rho}\gamma _{5}\rho \,,
\end{eqnarray}%
which scale with $\mathcal{\omega }^{0}$ and $\mathcal{\omega }$,
respectively. One could add arbitrary constants with appropriate
scale-weight but this would imply the presence of a cosmological constant
term in the bulk and would reproduce a sum of quadratic terms in {the} $\mathfrak{osp}\left( 4|1\right) $ covariant {super field-strengths} \cite{AD}.

An alternative approach is to add new gauge fields with upper scale-weight
than the present ones. The minimal content that can be added consists of a bosonic and a fermionic gauge field. In particular, we propose a new antisymmetric
bosonic gauge field $A^{ab}=-A^{ba}$ with scale-weight $\mathcal{\omega }%
^{2} $ and an additional fermionic gauge field $\chi $ with scale-weight $%
\mathcal{\omega }^{3/2}$. Naturally{,} one could consider additional bosonic
gauge fields with scale-weight $\mathcal{\omega }$ and $\mathcal{\omega }%
^{2} $ but, as we shall see, this will not be necessary to recover the
supersymmetry invariance in the boundary.

The only boundary contributions constructed by using $\left\{ \omega
^{ab},V^{a},A^{ab},\psi ,\chi \right\} $ that are compatible with parity,
Lorentz invariance and that do not involve a scaling parameter are given by
the following topological terms:%
\begin{eqnarray}
&&d\left( A^{ab}\mathcal{R}^{cd}+\omega _{\;f}^{a}\omega ^{fb}A^{cd}+2\omega
_{\;f}^{a}A^{fb}\omega ^{cd}+\omega ^{ab}\mathcal{F}^{cd}\right) \epsilon
_{abcd}=2\mathcal{R}^{ab}\mathcal{F}^{cd}\epsilon _{abcd}\,,  \notag \\
&&d\left( D\bar{\chi}\gamma _{5}\psi +D\bar{\psi}\gamma _{5}\chi \right) =2%
\bar{\sigma}\gamma _{5}\rho +\frac{1}{4}\mathcal{R}^{ab}\bar{\chi}\gamma
^{cd}\psi \epsilon _{abcd}\,,  \label{bdyt}
\end{eqnarray}%
where we have defined $\sigma \equiv D\chi $ and $\mathcal{F}^{ab}\equiv
DA^{ab}$ as the respective covariant {derivatives} of the new gauge fields.
Thus, the boundary Lagrangian reads
\begin{equation}
\mathcal{L}_{\text{bdy}}=\alpha \left( 2\bar{\sigma}\gamma _{5}\rho +\frac{1%
}{4}R^{ab}\bar{\chi}\gamma ^{cd}\psi \epsilon _{abcd}\right) +\beta \left( 2%
\mathcal{R}^{ab}\mathcal{F}^{cd}\epsilon _{abcd}\right) ,  \label{bdylagr}
\end{equation}%
where $\alpha $ and $\beta $ are constant parameters. Note that the boundary
Lagrangian has scale-weight $\mathcal{\omega }^{2}$ as the bulk Lagrangian.

Then, we have the following full Lagrangian{:}
\begin{equation}
\begin{split}
\mathcal{L}_{\text{full}}& =\mathcal{L}_{\text{bulk}}+\mathcal{L}_{\text{bdy}%
} \\
& =\epsilon _{abcd}\mathcal{R}^{ab}V^{c}V^{d}+4\bar{\psi}V^{a}\gamma
_{a}\gamma _{5}\rho \\
& \quad +\alpha \left( 2\bar{\sigma}\gamma _{5}\rho +\frac{1}{4}\mathcal{R}%
^{ab}\bar{\chi}\gamma ^{cd}\psi \epsilon _{abcd}\right) +\beta \left( 2%
\mathcal{R}^{ab}\mathcal{F}^{cd}\epsilon _{abcd}\right) \, .
\end{split}
\label{full}
\end{equation}
The supersymmetry invariance of the full Lagrangian requires%
\begin{equation}
\delta _{\epsilon }\mathcal{L}_{\text{full}}\equiv {l} _{\epsilon }\mathcal{%
L}_{\text{full}}=\imath _{\epsilon }d\mathcal{L}_{\text{full}}+d(\imath
_{\epsilon }\mathcal{L}_{\text{full}})=0 \,.  \label{ellf}
\end{equation}%
Naturally, the boundary terms (\ref{bdyt}) that we have introduced do not
affect the bulk and the supersymmetry invariance in the bulk is still
satisfied such that $\imath _{\epsilon }d\mathcal{L}_{\text{full}}=0$. Then,
the supersymmetry invariance of the full Lagrangian $\mathcal{L}_{\text{full}%
}$ requires to verify the condition $\imath _{\epsilon }\left( \mathcal{L}_{%
\text{full}}\right) |_{\partial \mathcal{M}}=0$.\ In particular, we have%
\begin{eqnarray}
\imath _{\epsilon }\left( \mathcal{L}_{\text{full}}\right) &=&\epsilon
_{abcd}\imath _{\epsilon }\left( \mathcal{R}^{ab}\right) V^{c}V^{d}+4\bar{%
\epsilon}V^{a}\gamma _{a}\gamma _{5}\rho +4\bar{\psi}V^{a}\gamma _{a}\gamma
_{5}\imath _{\epsilon }\left( \rho \right)  \notag \\
&&+\alpha \left\{ 2\imath _{\epsilon }\left( \bar{\sigma}\right) \gamma
_{5}\rho +2\bar{\sigma}\gamma _{5}\imath _{\epsilon }\left( \rho \right) +%
\frac{1}{4}\left[ \imath _{\epsilon }\left( \mathcal{R}^{ab}\right) \bar{%
\chi}\gamma ^{cd}\psi +\mathcal{R}^{ab}\bar{\chi}\gamma ^{cd}\epsilon
\right] \epsilon _{abcd}\right\}  \notag \\
&&+2\beta \left[ \imath _{\epsilon }\left( \mathcal{R}^{ab}\right) \mathcal{F%
}^{cd}\epsilon _{abcd}+\mathcal{R}^{ab}\imath _{\epsilon }\left( \mathcal{F}%
^{cd}\right) \epsilon _{abcd}\right] \,.
\end{eqnarray}%
In general, $\imath _{\epsilon }\mathcal{L}_{\text{full}}$ is not zero{;} however{,} its projection on the boundary should be zero. One can see that the
field equations acquire non-trivial boundary contributions coming not only
from the boundary Lagrangian but also from the bulk Lagrangian (from the
total differentials originating from partial integration), which implies the
following constraints on the boundary:
\begin{equation}
\left\{ \begin{aligned} \mathcal{R}^{ab} \vert_{\partial \mathcal{M}} = & \,
0 \,, \\ \mathcal{F}^{ab} \vert_{\partial \mathcal{M}} = & - \frac{1}{2\beta}
\left( V^a V^b + \frac{\alpha}{4} \bar{\chi} \gamma^{ab} \psi
\right)_{\partial \mathcal{M}} \,, \\ \rho \vert_{\partial \mathcal{M}} = & \,
0 \,, \\ \sigma \vert_{\partial \mathcal{M}} = & -\frac{2}{\alpha} \left( V^a
\gamma_a \psi \right)_{\partial \mathcal{M}} \,. \end{aligned}\right.
\label{eqbdy}
\end{equation}
{Thus, the super curvatures on the boundary result to be fixed to constant values in an enlarged anholonomic basis (indeed, on the boundary they are fixed in terms of not only the supervielbein, but also of the extra $1$-form field $\chi$). Nevertheless, as we will see in the sequel, their rheonomic parametrization results to be well defined in ordinary superspace.
Thus, the boundary values of the super curvatures in superspace are dynamically fixed by the field equations of the full Lagrangian (understood as bulk plus boundary contributions).}

Upon use of (\ref{eqbdy}) we find that
\begin{equation}
\imath _{\epsilon }\left( \mathcal{L}_{\text{full}}\right) |_{\partial
\mathcal{M}}=0\qquad \forall \alpha ,\beta  \, .
\end{equation}%
Thus the supersymmetry invariance of the full Lagrangian is restored in the
presence of a non-trivial boundary. The full Lagrangian can be then written
in terms of (\ref{eqbdy}) à la MacDowell-Mansouri \cite{MM} for $\alpha =4$
and $\beta =\frac{1}{2}$,%
\begin{equation}
{\mathcal{L}_{\text{full}}=\mathcal{R}^{ab} F
^{cd}\epsilon _{abcd} + 8 \bar{\Xi} \gamma _{5}\rho \,\,, }
\label{nMM}
\end{equation}%
{where we have defined}
\begin{eqnarray}
{F^{ab}} &{=}& {\mathcal{F}^{ab}+V^{a}V^{b}+\bar{\chi}\gamma ^{ab}\psi \,,} \label{FabMcurv} \\
{\Xi} &{=} & {\sigma +\frac{1}{2}\,V^{a}\gamma _{a}\,\psi \,.} \label{ChiMcurv}
\end{eqnarray}
It is important to point out that the supersymmetry invariance of the flat
supergravity Lagrangian on the boundary can be restored by adding new gauge
fields enlarging the Poincaré symmetry. In particular, this can be done for
arbitrary values of $\alpha $ and $\beta $. Nevertheless, the full
Lagrangian à la MacDowell-Mansouri appears for specific values of the
constants $\alpha $ and $\beta $ {($\alpha =4$
and $\beta =\frac{1}{2}$)}.

One could think that a dynamical Lagrangian can be recovered from the
original MacDowell-Mansouri Lagrangian constructed from the $OSp\left(
4|1\right) $ covariant super curvatures. Although the bulk Lagrangians can
be related through a flat limit, the vanishing cosmological constant limit
cannot be naively applied in presence of a non-trivial boundary. Indeed, in
the flat limit, the only term that remains is a boundary topological term and
corresponds to a Gauss-Bonnet term. This is mainly due to the presence of
the $\ell $ parameter {(length scale)} in every term of the bulk Lagrangian of the $OSp\left(
4|1\right) $ supergroup. Thus, in order to obtain the pure supergravity
action without cosmological constant term using the MacDowell-Mansouri
formalism, it is necessary to enlarge the symmetry. This enlargement, as we
have shown, does not modify the bulk Lagrangian but affects only the boundary
allowing to restore the supersymmetry invariance.

{Let us observe that, interestingly, for $\alpha =4$ and $\beta =\frac{1}{2}$ we also find out} that the curvatures (\ref%
{eqbdy}) reproduce the minimal Maxwell covariant super curvatures \cite{BGKL, CRS}{, namely \eqref{FabMcurv}, \eqref{ChiMcurv}, and}
\begin{eqnarray}
R^{ab} &=&\mathcal{R}^{ab}\,,  \notag \\
\Psi &=&\rho \,,  \label{Mcurv} \\
R^{a} &=&DV^{a}-\frac{1}{2}\,\bar{\psi}\gamma ^{a}\psi \,,  \notag
\end{eqnarray}%
{which satisfy the Bianchi identities}
\begin{eqnarray}
DR^{ab} &=&0\,,  \notag \\
DF^{ab} &=&2R_{\text{ }c}^{a}A^{cb}+2R^{a}V^{b}+\bar{\Xi}\gamma ^{ab}\psi -%
\bar{\chi}\gamma ^{ab}\Psi \,,  \notag \\
D\Psi &=&\frac{1}{4}\,R^{ab}\gamma _{ab}\psi \,,  \label{BI} \\
D\Xi &=&\frac{1}{4}\,R^{ab}\gamma _{ab}\chi +\frac{1}{2}\,R^{a}\gamma
_{a}\psi -\frac{1}{2}\,V^{a}\gamma _{a}\Psi \,,  \notag \\
DR^{a} &=&R_{\text{ }b}^{a}V^{b}+\bar{\psi}\gamma ^{a}\Psi \,.  \notag
\end{eqnarray}%
In particular, the {super-Maxwell curvatures vanish at the boundary:}
\begin{eqnarray}
R^{ab}|_{\partial \mathcal{M}} &=&0\,,  \notag \\
F^{ab}|_{\partial \mathcal{M}} &=&0\,,  \notag \\
\Psi |_{\partial \mathcal{M}} &=&0\,, \\
\Xi |_{\partial \mathcal{M}} &=&0\,.  \notag
\end{eqnarray}%
{Then, the full MacDowell-Mansouri like Lagrangian \eqref{nMM} can be rewritten as
\begin{equation}
{\mathcal{L}_{\text{full}}=R^{ab} F
^{cd}\epsilon _{abcd} + 8 \bar{\Xi} \gamma _{5}\Psi \,\, }
\end{equation}
in terms of the Maxwell super curvatures \eqref{FabMcurv}, \eqref{ChiMcurv}, and \eqref{Mcurv}.}

Let us observe that the full Lagrangian obtained here cannot be directly
obtained as a flat limit of the $OSp\left( 4|1\right) $ one \cite{AD} due to
the presence of new gauge fields. Nevertheless, one could apply our approach
to the supergravity Lagrangian in presence of the cosmological constant {(and thus of a length scale $\ell$)} by
adding the extra gauge fields not only to the boundary Lagrangian but also
to the bulk Lagrangian{, such that the flat limit $\ell \rightarrow \infty $} reproduces the full Lagrangian (\ref{nMM}) obtained here.

Before studying the supersymmetry invariance of a supergravity theory with a
generalized cosmological constant in the presence of a {non-trivial} boundary, {let us provide the rheonomic parametrization} of the Maxwell super curvatures and the
supersymmetry transformation laws.

\subsection{Maxwell supersymmetry}

The minimal Maxwell superalgebra has been introduced in \cite{BGKL} in order
to describe a generalized four-dimensional\ superspace in the presence of a
constant abelian supersymmetric {field-strength} background. Such superalgebra
has the particularity to extend the Maxwell symmetry $\left\{
J_{ab},P_{a},Z_{ab}\right\} $ by adding two spinorial generators $\left\{
Q_{\alpha },\Sigma _{\alpha }\right\} $. In particular, the generators
satisfy the following non-vanishing {(anti)commutation relations:}
\begin{eqnarray}
\left[ J_{ab},J_{cd}\right] &=&\eta _{bc}J_{ad}-\eta _{ac}J_{bd}-\eta
_{bd}J_{ac}+\eta _{ad}J_{bc}\,,  \notag \\
\left[ J_{ab},P_{c}\right] &=&\eta _{bc}P_{a}-\eta _{ac}P_{b}\,,  \notag \\
\left[ P_{a},P_{b}\right] &=&Z_{ab}\,,  \notag \\
\left[ J_{ab},Z_{cd}\right] &=&\eta _{bc}Z_{ad}-\eta _{ac}Z_{bd}-\eta
_{bd}Z_{ac}+\eta _{ad}Z_{bc}\,,  \notag \\
\left[ J_{ab,}Q_{\alpha }\right] &=&-\frac{1}{2}\,\left( \gamma
_{ab}Q\right) _{\alpha }\,,  \label{sM} \\
\left[ J_{ab},\Sigma _{\alpha }\right] &=&-\frac{1}{2}\,\left( \gamma
_{ab}\Sigma \right) _{\alpha }\,,  \notag \\
\left[ P_{a},Q_{\alpha }\right] &=&-\frac{1}{2}\,\left( \gamma _{a}\Sigma
\right) _{\alpha }\,,  \notag \\
\left\{ Q_{\alpha },Q_{\beta }\right\} &=&\left( \gamma ^{a}C\right)
_{\alpha \beta }P_{a}\,,  \notag \\
\left\{ Q_{\alpha },\Sigma _{\beta }\right\} &=&-\frac{1}{2}\left( \gamma
^{ab}C\right) _{\alpha \beta }Z_{ab}\,.  \notag
\end{eqnarray}%

{Concerning} the {purely} bosonic level, the Maxwell algebra has been introduced in \cite{BCR,
Schrader}. Such symmetry and its generalizations have been recently useful
to recover General Relativity from Chern-Simons (CS) and Born-Infeld (BI)
gravity formalisms \cite{EHTZ, GRCS, CPRS1, CPRS2, CPRS3}. More recently,
there has been a new interest in exploring the three-dimensional Maxwell CS
gravity \cite{SSV, HR, CCFRS, AFGHZ, CMMRSV}. At the supersymmetric level,
the {super}-Maxwell family also appears in three spacetime dimensions allowing
to reproduce CS supergravity models \cite{CFRS, CFR, CPR}. Further
generalizations of the minimal Maxwell superalgebra can be found in \cite%
{AILW, AI, CR1, CR2, PR, Ravera} with diverse applications.

In the geometric approach, the most general ansatz for the super-Maxwell curvatures in the
{supervielbein} basis $\left\{ V^{a},\psi \right\} $ of superspace is given
by
\begin{align}
R^{ab}& =R_{\;\;\;\;cd}^{ab}V^{c}\wedge V^{d}+\bar{\Theta}%
_{\;\;\;\;c}^{ab}\psi \wedge V^{c}+ {\lambda_1} \bar{\psi}\wedge \gamma ^{ab}\psi
\,,  \notag \\
R^{a}& =R_{\;\;bc}^{a}V^{b}\wedge V^{c}+\bar{\Theta}_{\;\;b}^{a}\psi \wedge
V^{b}+ {\lambda_2} \bar{\psi}\wedge \gamma ^{a}\psi \,,  \notag \\
F^{ab}& =F_{\;\;\;\;cd}^{ab}V^{c}\wedge V^{d}+\bar{\Lambda}%
_{\;\;\;\;c}^{ab}\psi \wedge V^{c}+ {\lambda_3} \bar{\psi}\wedge \gamma ^{ab}\psi
\,,  \label{sMbasis} \\
{\Psi} & = {\Psi} _{ab}V^{a}\wedge V^{b}+ {\lambda_4} \gamma _{a}\psi \wedge
V^{a}+\Omega _{\alpha \beta }\psi ^{\alpha }\wedge \psi ^{\beta }\,,  \notag
\\
\Xi & =\Xi _{ab}V^{a}\wedge V^{b}+ {\lambda_5} \gamma _{a}\psi \wedge V^{a}+\tilde{%
\Omega}_{\alpha \beta }\psi ^{\alpha }\wedge \psi ^{\beta }\, {,} \notag
\end{align}%
{where $\bar{\Theta}_{\;\;\;\;c}^{ab}$, $\bar{\Theta}_{\;\;b}^{a}$, $\bar{\Lambda}
_{\;\;\;\;c}^{ab}$ are spinor-tensors, $\Omega _{\alpha \beta }$, $\tilde{\Omega}_{\alpha \beta }$ are Majorana spinor-valued matrices, and the} $\lambda_i$'s ($i=1,2,\ldots,5$) constant parameters. Considering the on-shell condition $R^{a}=0$ and {studying the various sectors of the} on-shell Bianchi
identities (\ref{BI}) in superspace, one can show that the full set of
super curvatures are parameterized as follows{:}
\begin{align}
R^{ab}& =R_{\;\;\;\;cd}^{ab}V^{c}V^{d}+\bar{\Theta}_{\;\;\;\;c}^{ab}\psi
V^{c}\,,  \notag \\
R^{a}& =0\,,  \notag \\
F^{ab}& =F_{\;\;\;\;cd}^{ab}V^{c}V^{d}+\bar{\Lambda}_{\;\;\;\;c}^{ab}\psi
V^{c}\,,  \label{sMpara} \\
{\Psi} & = {\Psi} _{ab}V^{a}V^{b}\,,  \notag \\
\Xi & ={\lambda_5} \gamma _{a}\psi V^{a}\,,  \notag
\end{align}%
with%
\begin{equation}
\bar{\Theta}_{\;\;\;\;c}^{ab}=\epsilon ^{abde}\left( \bar{{\Psi}}_{cd}\gamma
_{e}\gamma _{5}+\bar{{\Psi}}_{ec}\gamma _{d}\gamma _{5}-\bar{{\Psi}}_{de}\gamma
_{c}\gamma _{5}\right) \, .
\end{equation}

Such parametrization allows us to obtain explicitly the supersymmetry
transformation laws. In particular, in the rheonomic approach, a {1-form superfield transformation on spacetime} takes the form%
\begin{equation*}
\delta \mu ^{A}=\left( \nabla \epsilon \right) ^{A}+\imath _{\epsilon
}{R}^{A}\,,
\end{equation*}%
where $\epsilon ^{A}$ {are the gauge parameters} and ${R}^{A}$ {are the parameterized super curvatures}. Then, restricting us to supersymmetry transformation and
considering the parametrization (\ref{sMpara}) of the super-Maxwell
curvatures, we find the following supersymmetry transformation laws:%
\begin{align}
\delta _{\epsilon }\omega ^{ab}& =2\bar{\Theta}_{\;\;\;\;c}^{ab}\epsilon
V^{c}\,,  \notag \\
\delta _{\epsilon }V^{a}& =\bar{\epsilon}\gamma ^{a}\psi \,,  \notag \\
\delta _{\epsilon }{A^{ab}}& =\bar{\xi}\gamma ^{ab}\epsilon +2\bar{\Lambda}%
_{\;\;\;\;c}^{ab}\epsilon V^{c}\,, \\
\delta _{\epsilon }\psi & =D\epsilon \,,  \notag \\
\delta _{\epsilon }{\chi} & =V^{a}\gamma _{a}\epsilon \,,  \notag
\end{align}%
where we have set ${\lambda_5} =-\frac{1}{2}$.

\section{Supergravity with generalized cosmological constant in presence of a non-trivial boundary}\label{Section4}

The presence of additional bosonic generators has been recently related to a
generalization of the cosmological constant \cite{AKL, CRS, CIRR, PR2}.
Nevertheless, as we have {previously seen in Section \ref{Section3}}, the introduction {of additional gauge fields} does not necessary imply the presence of a
cosmological constant term in a (super)gravity action. In order to include {a} cosmological constant contribution to our supergravity model, it is
necessary to switch on an explicit scale. In particular, we can write the
following super curvatures:\footnote{%
Notice that these super curvatures can be obtained by considering the same scale-weight for the gauge fields as in the flat case. The main difference with the previous curvatures consists in the explicit presence of the length parameter $\ell$.}
\begin{align}
\mathfrak{R}^{ab}& \equiv d\omega ^{ab}+\omega _{\;c}^{a}\omega ^{cb}+\frac{1%
}{2\ell }\bar{\psi}\gamma ^{ab}\psi \,,  \notag \\
\mathfrak{R}^{a}& \equiv DV^{a}+\frac{1}{\ell ^{2}}A_{\;b}^{a}V^{b}-\frac{1}{%
2}\bar{\psi}\gamma ^{a}\psi -\frac{1}{\ell }\bar{\psi}\gamma ^{a}\chi -\frac{%
1}{2\ell ^{2}}\bar{\chi}\gamma ^{a}\chi \,,  \notag \\
\rho & \equiv D\psi \,,  \label{sc1} \\
\mathfrak{F}^{ab}& \equiv DA^{ab}+V^{a}V^{b}+\bar{\chi}\gamma ^{ab}\psi +%
\frac{1}{\ell ^{2}}A_{\;c}^{a}A^{cb}+\frac{1}{2\ell }\bar{\chi}\gamma
^{ab}\chi \, ,  \notag \\
\Omega & \equiv D\chi +\frac{1}{2}V^{a}\gamma _{a}\psi +\frac{1}{2\ell }%
V^{a}\gamma _{a}\chi +\frac{1}{4\ell }A^{ab}\gamma _{ab}\psi +\frac{1}{4\ell
^{2}}A^{ab}\gamma _{ab}\chi \,,  \notag
\end{align}%
being, as usual, $D=d+\omega $ the Lorentz covariant derivative.
The next step is the explicit construction of the bulk Lagrangian.

\subsection{Rheonomic construction of the bulk Lagrangian}

The most general ansatz for the geometric bulk Lagrangian can be written as
\begin{equation}
\mathcal{L}=\mu ^{(4)}+R^{A}\mu _{A}^{(2)}+R^{A}R^{B}\mu _{AB}^{(0)} \, ,
\label{ansatzAdSsM}
\end{equation}%
where the upper index $\left( p\right) $ denotes the degree of the related $%
p $-forms. Here, the $R^{A}$'s are the super curvatures defined by eq. (\ref%
{sc1}), which are invariant under the following rescaling{:}%
\begin{equation}
\omega ^{ab}\rightarrow \omega ^{ab}\,,\quad V^{a}\rightarrow \omega
V^{a}\,,\quad A^{ab}\rightarrow \omega ^{2}A^{ab}\,,\quad \psi \rightarrow
\omega ^{1/2}\psi \,,\quad \chi \rightarrow \omega ^{3/2}\chi \,.
\end{equation}%
In particular, the Lagrangian scales with $\omega ^{2}$, being $\omega ^{2}$
the scale-weight of the EH term. On the other hand, since we are interested
in the construction of the bulk Lagrangian, we can set $R^{A}R^{B}\mu
_{AB}^{(0)}=0${, since terms of this type correspond to boundary contributions.} Then, considering
the explicit form of the super curvatures (\ref{sc1}) and applying the
parity conservation law, we are left with the following Lagrangian:%
\begin{eqnarray}
\mathcal{L} &=&\epsilon _{abcd}\mathfrak{R}^{ab}V^{c}V^{d}+\frac{\alpha _{1}%
}{\ell ^{2}}\,\epsilon _{abcd}\mathfrak{F}^{ab}V^{c}V^{d}+\alpha _{2}\bar{%
\psi}\gamma _{a}\gamma _{5}\rho V^{a}+\frac{\alpha _{3}}{\ell }\,\bar{\psi}%
\gamma _{a}\gamma _{5}\Omega V^{a}  \notag \\
&&+\frac{\alpha _{4}}{\ell ^{2}}\,\bar{\chi}\gamma _{a}\gamma _{5}\Omega
V^{a}+\frac{\alpha _{5}}{\ell }\,\bar{\chi}\gamma _{a}\gamma _{5}\rho V^{a}+%
\frac{\alpha _{6}}{\ell ^{2}}\,\epsilon _{abcd}V^{a}V^{b}V^{c}V^{d}+\frac{%
\alpha _{7}}{\ell }\,\epsilon _{abcd}\bar{\psi}\gamma ^{ab}\psi V^{c}V^{d}
\label{Lbulk} \\
&&+\frac{\alpha _{8}}{\ell ^{2}}\,\epsilon _{abcd}\,\bar{\psi}\gamma
^{ab}\chi V^{c}V^{d}+\frac{\alpha _{9}}{\ell ^{3}}\,\epsilon _{abcd}\bar{\chi%
}\gamma ^{ab}\chi V^{c}V^{d}\,,  \notag
\end{eqnarray}%
where we have set the coefficient of the first term in \eqref{Lbulk} to $1$.
The $\alpha _{i}$'s constants are dimensionless parameters which can
be determined from the equations of motion. Indeed, considering the
variation of the Lagrangian with respect to the gauge fields we obtain the
following field equations:
\begin{align}
\delta _{\omega }\mathcal{L}=\delta _{A}\mathcal{L}=0& \rightarrow \epsilon
_{abcd}\mathfrak{R}^{c}V^{d}=0\,,  \notag \\
\delta _{V}\mathcal{L}=0& \rightarrow 2\epsilon _{abcd}\left( \mathfrak{R}%
^{ab}+ {\frac{1}{\ell^2}} \mathfrak{F}^{ab}\right) V^{c}+4\left( \bar{\psi}+\frac{1}{\ell }\chi
\right) \gamma _{d}\gamma _{5}\rho +\frac{4}{\ell }\left( \bar{\psi}+\frac{1%
}{\ell }\bar{\chi}\right) \gamma _{d}\gamma _{5}\Omega =0\,,  \label{eom} \\
\delta _{\psi }\mathcal{L}=\delta _{\chi }\mathcal{L}=0& \rightarrow
8V^{a}\gamma _{a}\gamma _{5}\left( \rho +\frac{1}{\ell }\Omega \right)
+4\gamma _{a}\gamma _{5}\left( \psi +\frac{1}{\ell }\chi \right) \mathfrak{R}%
^{a} {=0} \,,  \notag
\end{align}%
where, in order to recover consistent field equations involving the (generalized) curvature two-forms (\ref{sc1}), we have set
\begin{align}
& \alpha _{1} \, {=} \, 1\,,  \notag \\
& \alpha _{2} \, {=} \, \alpha _{3} \, = \, \alpha _{4} \, = \, \alpha _{5}= \, 4\,,  \notag \\
& \alpha _{6} \, {=} \, \alpha _{7} \, = \, \alpha _{9} \, =\, -\frac{1}{2}\,, \\
& \alpha _{8} \, {=} \, -1\,.  \notag
\end{align}%
{In particular, setting $\alpha_{1}=1$ and $\alpha_{2}=\alpha_{3}=\alpha_{4}=\alpha_{5}=4$, $\delta_{\omega}\mathcal{L}=0$ reproduces the field equation for the generalized supertorsion, that is $\epsilon_{abcd}\mathfrak{R}^{c}V^{d}=0\,$ (thus, the first equation corresponds to the
vanishing of the generalized supertorsion $\mathfrak{R}^{a}$). The variation of the Lagrangian with respect to the gauge field $A^{ab}$ yields the same result. On the other hand, the curvatures $\mathfrak{R}^{ab}$ and $\mathfrak{F}^{ab}$ appear explicitly in the generalized equation of motion obtained when varying the Lagrangian with respect to the vielbein $V^{a}$ by setting $\alpha_{6}=\alpha_{7}=\alpha_{9}=-\frac{1}{2}$ and $\alpha_{8}=-1$. Finally, from the variation of the Lagrangian with respect to the gravitino $\psi$ (as well as from the variation with respect to $\chi$) one obtains the generalized equation involving the curvature $2$-forms $\rho$ and $\Omega$ (together with the generalized supertorsion $\mathfrak{R}^{a}$).}
Let us observe that the equations of motion (\ref{eom}) are a generalization of the $OSp(4|1)$ ones, the only differences being related to the presence of the new fields $A^{ab}$ and $\chi$ (and of their respective super field-strength).
{Moreover, the equations of motion obtained by varying the Lagrangian with respect to the new fields $A^{ab}$ and $\chi$ are just a double of the field equations obtained from the variation with respect to the spin-connection $\omega^{ab}$ and the gravitino $\psi$, respectively.}
With {these} particular values of the constants, the equations of motion of the
Lagrangian admit an AdS vacuum solution with a generalized cosmological
constant {given by $\frac{1}{\ell^2}\epsilon _{abcd}\mathfrak{F}^{ab}V^{c}V^{d}-\frac{1}{2\ell^2}\epsilon_{abcd}V^{a}V^{b}V^{c}V^{d}$, where $\mathfrak{F}^{ab}$ has been defined in \eqref{sc1}. As was noticed in \cite{AKL}, the presence of an additional gauge field in gravity allows to define a generalized cosmological constant term written in terms of the vielbein $V^{a}$ and the new gauge field $A^{ab}$. Such feature was then extended to supergravity in \cite{CRS, CIRR, PR2}, offering, in particular, an alternative way of introducing a generalized cosmological term in this context. It is interesting to note that flat supergravity with boundary could appear from a supergravity action in presence of a more general cosmological term.}

Thus, the bulk Lagrangian can be rewritten in terms of Lorentz-type
curvatures as follows{:}%
\begin{eqnarray}
\mathcal{L}_{\text{bulk}} &=&\epsilon _{abcd}R^{ab}V^{c}V^{d\ }+\frac{1}{%
\ell ^{2}}\,\epsilon _{abcd}\mathbb{F}^{ab}V^{c}V^{d}+4\bar{\psi}\gamma
_{a}\gamma _{5}\rho V^{a}+\frac{4}{\ell }\,\bar{\psi}\gamma _{a}\gamma
_{5}\Phi V^{a}  \notag \\
&&+\frac{4}{\ell ^{2}}\,\bar{\chi}\gamma _{a}\gamma _{5}\Phi V^{a}+\frac{4}{%
\ell }\,\bar{\chi}\gamma _{a}\gamma _{5}\rho V^{a}+\frac{1}{2\ell ^{2}}%
\,\epsilon _{abcd}V^{a}V^{b}V^{c}V^{d}  \label{bulkL} \\
&&+\frac{1}{\ell }\,\epsilon _{abcd}\bar{\psi}\gamma ^{ab}\psi V^{c}V^{d}+%
\frac{2}{\ell ^{2}}\,\epsilon _{abcd}\bar{\chi}\gamma ^{ab}\psi V^{c}V^{d}+%
\frac{1}{\ell ^{3}}\,\epsilon _{abcd}\bar{\chi}\gamma ^{ab}\chi V^{c}V^{d}\,,
\notag
\end{eqnarray}%
with
\begin{align}
R^{ab}& =d\omega ^{ab}+\omega _{\;c}^{a}\omega ^{cb}\,,  \notag \\
\mathfrak{R}^{a}& =DV^{a}+\frac{1}{\ell ^{2}}A_{\;b}^{a}V^{b}-\frac{1}{2}%
\bar{\psi}\gamma ^{a}\psi -\frac{1}{\ell }\bar{\psi}\gamma ^{a}\chi -\frac{1%
}{2\ell ^{2}}\bar{\chi}\gamma ^{a}\chi \,,  \notag \\
\rho & =D\psi \,,  \label{Ltc} \\
\mathbb{F}^{ab}& =DA^{ab}+\frac{1}{\ell ^{2}}A_{\;c}^{a}A^{cb}\,,  \notag \\
\Phi & =D\chi +\frac{1}{4\ell }A^{ab}\gamma _{ab}\psi +\frac{1}{4\ell ^{2}}%
A^{ab}\gamma _{ab}\chi \,.  \notag
\end{align}%
Let us note that the presence of the length parameter $\ell $ in the super
curvatures allows to introduce, in an alternative way, a generalized
supersymmetric cosmological constant term in the Lagrangian. Such Lagrangian
can be seen as a deformation of the usual supergravity Lagrangian for the $%
\mathfrak{osp}(4|1)$ superalgebra. Interestingly, as {for the case of} the $\mathfrak{osp}(4|1) $ supergravity, the vanishing cosmological constant limit $\ell
\rightarrow \infty $ leads to the flat supergravity Lagrangian. However, as
we have discussed in {Section \ref{Section3}}, the supersymmetry invariance {of flat} supergravity on a manifold with boundary is recovered by adding non-trivial {boundary (topological) terms.} Then, it seems that the flat limit in presence of a
boundary requires to consider a deformation of the full Lagrangian for the $%
\mathfrak{osp}(4|1)$ superalgebra. Thus, the bulk Lagrangian (\ref{bulkL})
obtained here seems a good candidate to consider in presence of a boundary.
{Observe that, when an explicit scale is switched on, terms involving the new $1$-form fields $A^{ab}$ and $\chi$ are allowed to appear in the bulk Lagrangian, which was not the case for flat supergravity, where, however, as we have previously shown, they play a crucial role in restoring the supersymmetry invariance in the presence of a non-trivial boundary.}

Before studying the explicit boundary contributions required to assure
supersymmetry invariance we shall {provide} the supersymmetry
transformation laws under which the bulk Lagrangian (\ref{bulkL}) is
invariant.

The supersymmetry transformations can be obtained from the {rheonomic} parametrization of the super curvatures which are determined from the study of the Bianchi identities of
the Lorentz-type curvatures (\ref{Ltc}). In particular, the super curvatures
(\ref{Ltc}) fulfill the following Bianchi identities:
\begin{align}
DR^{ab}& =0\,,  \notag \\
D\mathfrak{R}^{a}& =R_{\;b}^{a}V^{b}+\frac{1}{\ell ^{2}}\mathbb{F}%
_{\;b}^{a}V^{a}-\frac{1}{\ell ^{2}}A_{\;b}^{a}\mathfrak{R}^{a}+\bar{\psi}%
\gamma ^{a}\rho +\frac{1}{\ell }\bar{\chi}\gamma ^{a}\rho +\frac{1}{\ell }%
\bar{\psi}\gamma ^{a}\Phi +\frac{1}{\ell ^{2}}\bar{\chi}\gamma ^{a}\Phi \,,
\notag \\
D\rho & =\frac{1}{4}R^{ab}\gamma _{ab}\psi \,,  \label{llbi} \\
D\mathbb{F}^{ab}& =2R_{\;c}^{a}A^{cb}+\frac{2}{\ell ^{2}}\mathbb{F}%
_{\;c}^{a}A^{cb}\,,  \notag \\
D\Phi & =\frac{1}{4}R^{ab}\gamma _{ab}\chi +\frac{1}{4\ell }\mathbb{F}%
^{ab}\gamma _{ab}\psi -\frac{1}{4\ell }A^{ab}\gamma _{ab}\rho +\frac{1}{%
4\ell ^{2}}\mathbb{F}^{ab}\gamma _{ab}\chi -\frac{1}{4\ell ^{2}}A^{ab}\gamma
_{ab}\Phi \,.  \notag
\end{align}%
Notice that the {super curvatures} above are defined in a superspace larger
than the ordinary one. In the following, we will ask the curvatures
parametrization to be well defined in ordinary superspace by exploiting the
rheonomic approach. One can show that the Bianchi identities (\ref{llbi})
are satisfied by parameterizing (on-shell) the {super curvatures} in the
following way:
\begin{align}
R^{ab}& = {\mathcal{R}}_{\;\;\;\;cd}^{ab}V^{c}V^{d}+{\bar{\Pi}}_{\;\;\;\;c}^{ab}\psi
V^{c}+\varsigma _{1}\bar{\psi}\gamma ^{ab}\psi \,,  \notag \\
\mathfrak{R}^{a}& =0\,,  \notag \\
\mathbb{F}^{ab}& = {\mathcal{F}}_{\;\;\;\;cd}^{ab}V^{c}V^{d}+{\bar{\Delta}}_{\;\;\;\;c}^{ab}\psi V^{c}\,, \label{rhparell} \\
\rho & =\rho _{ab}V^{a}V^{b}+{\varsigma _{2}}\gamma _{a}\psi V^{a}\,,  \notag
\\
\Phi & =\Phi _{ab}V^{a}V^{b}+{\varsigma _{3}}\gamma _{a}\psi V^{a}\,,  \notag
\end{align}%
where
\begin{align}
& {\bar{\Pi}}_{\;\;\;\;c}^{ab}=\epsilon ^{abde}\left( \bar{\rho}_{cd}\gamma
_{e}\gamma _{5}+\bar{\rho}_{ec}\gamma _{d}\gamma _{5}-\bar{\rho}_{de}\gamma
_{c}\gamma _{5}\right) \,,  \notag \\
& {\bar{\Delta}}_{\;\;\;\;c}^{ab}=\epsilon ^{abde}\left( \bar{\Phi}%
_{cd}\gamma _{e}\gamma _{5}+\bar{\Phi}_{ec}\gamma _{d}\gamma _{5}-\bar{\Phi}%
_{de}\gamma _{c}\gamma _{5}\right) \,, \\
& \varsigma _{1}=-{\varsigma _{2}}-\frac{1}{\ell }{\varsigma _{3}} \,,  \notag
\end{align}%
and where we have set $\mathfrak{R}^{a}=0$ (on-shell condition). Such
parametrization of the super curvatures in ordinary superspace provides us
with the supersymmetry transformation laws for the $1$-form gauge fields,
which read as follows:
\begin{align}
\delta _{\epsilon }\omega ^{ab}& =2{\bar{\Pi}}_{\;\;\;\;c}^{ab}\epsilon
V^{c}+2\varsigma _{1}\bar{\epsilon}\gamma ^{ab}\psi \,,  \notag \\
\delta _{\epsilon }V^{a}& =\bar{\epsilon}\gamma ^{a}\psi -\frac{1}{\ell }%
\bar{\chi}\gamma ^{a}\epsilon \,,  \notag \\
\delta _{\epsilon }A^{ab}& =2 {\bar{\Delta}}_{\;\;\;\;c}^{ab}\epsilon V^{c}\,,
\label{susyt} \\
\delta _{\epsilon }\psi & =D\epsilon + {\varsigma _{2}} \gamma _{a}\epsilon
V^{a}\,,  \notag \\
\delta _{\epsilon }\chi & = {\varsigma _{3}}\gamma _{a}\epsilon V^{a}\,.  \notag
\end{align}

It is important to point out that the spacetime Lagrangian (\ref{bulkL}) is
invariant up to boundary terms under the supersymmetry transformations (\ref%
{susyt}) of the gauge fields on spacetime. If the spacetime background has a
non-trivial boundary, we have to check explicitly the supersymmetry
invariance.

\subsection{Supersymmetry invariance in presence of a boundary}

Let us now consider the supergravity theory previously introduced in
presence of a non-trivial spacetime boundary, and let us study the
supersymmetry invariance of it. In particular, we shall see that appropriate
boundary terms are required in order to restore the supersymmetry invariance
of the full Lagrangian {given by} the bulk Lagrangian (\ref%
{bulkL}) plus boundary contributions. {Although the supersymmetry invariance
in the bulk is satisfied, the invariance of the Lagrangian when a
boundary is present, is not trivially satisfied:}
\begin{equation}
\imath _{\epsilon }\mathcal{L}_{\text{bulk}}|_{\partial \mathcal{M}}\neq 0\,.
\end{equation}

Then, for recovering the supersymmetry invariance of the theory, it is
necessary to modify the bulk Lagrangian by adding suitable boundary terms.
The only possible boundary contributions compatible with parity and Lorentz
invariance are given by
\begin{align}
& d\left( \omega ^{ab}R^{cd}+\omega _{\;f}^{a}\omega ^{fb}\omega
^{cd}\right) \epsilon _{abcd}=\epsilon _{abcd}R^{ab}R^{cd}\,,  \notag \\
& d\left( A^{ab}R^{cd}+\omega _{\;f}^{a}\omega ^{fb}A^{cd}+2\omega
_{\;f}^{a}A^{fb}\omega ^{cd}+\omega ^{ab}\mathbb{F}^{cd}+\frac{1}{\ell ^{2}}%
A_{\;f}^{a}A^{fb}\omega ^{cd}+\frac{2}{\ell ^{2}}\omega
_{\;f}^{a}A^{fb}A^{cd}\right.  \notag \\
& \left. \frac{1}{\ell ^{2}}A^{ab}\mathbb{F}^{cd}+\frac{1}{\ell ^{4}}%
A_{\;f}^{a}A^{fb}A^{cd}\right) \epsilon _{abcd}=\epsilon _{abcd}\left(
2R^{ab}\mathbb{F}^{cd}+\frac{1}{\ell ^{2}}\mathbb{F}^{ab}\mathbb{F}%
^{cd}\right) \,,  \notag \\
& d\left( \bar{\psi}\gamma _{5}\rho \right) =\bar{\rho}\gamma _{5}\rho +%
\frac{1}{8}R^{ab}\bar{\psi}\gamma ^{cd}\psi \epsilon _{abcd}\,,
\label{bdyt2} \\
& d\left( \bar{\psi}\gamma _{5}\Phi +\bar{\chi}\gamma _{5}\rho +\frac{1}{%
\ell }\bar{\chi}\gamma _{5}\Phi \right) =2\bar{\rho}\gamma _{5}\Phi +\bar{%
\Phi}\gamma _{5}\Phi +\frac{1}{4}R^{ab}\bar{\psi}\gamma ^{cd}\chi \epsilon
_{abcd}+\frac{1}{8\ell }\mathbb{F}^{ab}\bar{\psi}\gamma ^{cd}\psi \epsilon
_{abcd}  \notag \\
& +\frac{1}{4\ell ^{2}}\mathbb{F}^{ab}\bar{\psi}\gamma ^{cd}\chi \epsilon
_{abcd}+\frac{1}{8\ell }R^{ab}\bar{\chi}\gamma ^{cd}\chi \epsilon _{abcd}+%
\frac{1}{8\ell ^{3}}\mathbb{F}^{ab}\bar{\chi}\gamma ^{cd}\chi \epsilon
_{abcd}\,.  \notag
\end{align}%
The boundary terms (\ref{bdyt2}) allow us to write the following boundary
Lagrangian{:}
\begin{eqnarray}
\mathcal{L}_{\text{bdy}} &=&\lambda \epsilon _{abcd}R^{ab}R^{cd}+\pi \left(
\bar{\rho}\gamma _{5}\rho +\frac{1}{8}R^{ab}\bar{\psi}\gamma ^{cd}\psi
\epsilon _{abcd}\,\right) +\mu \epsilon _{abcd}\left( 2R^{ab}\mathbb{F}^{cd}+%
\frac{1}{\ell ^{2}}\mathbb{F}^{ab}\mathbb{F}^{cd}\right)  \notag \\
&&+\nu \left( 2\bar{\rho}\gamma _{5}\Phi +\bar{\Phi}\gamma _{5}\Phi +\frac{1%
}{4}R^{ab}\bar{\psi}\gamma ^{cd}\chi \epsilon _{abcd}+\frac{1}{8\ell }%
\mathbb{F}^{ab}\bar{\psi}\gamma ^{cd}\psi \epsilon _{abcd}\right.  \notag \\
&&\left. +\frac{1}{4\ell ^{2}}\mathbb{F}^{ab}\bar{\psi}\gamma ^{cd}\chi
\epsilon _{abcd}+\frac{1}{8\ell }R^{ab}\bar{\chi}\gamma ^{cd}\chi \epsilon
_{abcd}+\frac{1}{8\ell ^{3}}\mathbb{F}^{ab}\bar{\chi}\gamma ^{cd}\chi
\epsilon _{abcd}\right) \,,  \label{bdyLagrangian}
\end{eqnarray}%
where $\lambda ,\pi ,\mu ,\nu $ are constant parameters. In order to have a
consistently defined limit $\ell \rightarrow \infty $ (flat limit) at the
level of the full Lagrangian, one should drop out the terms involving
positive powers of $\ell $. From the Lagrangian (\ref{bdyLagrangian}) we see
that all terms have scale-weight $\mathcal{\omega }^{2}$ {except those
proportional to $\lambda$ and $\pi$.} Thus, {in order to have} an {appropriately} scaled boundary Lagrangian one should define new {(dimensionless)} constants{, $\lambda^{\prime}$ and $\pi^{\prime}$,} such that $\lambda =\ell
^{2}\lambda ^{\prime }$ and $\pi =\ell \pi ^{\prime }$, which implies
positive powers of $\ell$. Then, the terms proportional to $\lambda$ and $\pi$ must be dropped out.

In this way, we are left with%
\begin{eqnarray}
\mathcal{L}_{\text{bdy}} &=&\mu \epsilon _{abcd}\left( 2R^{ab}\mathbb{F}%
^{cd}+\frac{1}{\ell ^{2}}\mathbb{F}^{ab}\mathbb{F}^{cd}\right)  \notag \\
&&+\nu \left( 2\bar{\rho}\gamma _{5}\Phi +\bar{\Phi}\gamma _{5}\Phi +\frac{1%
}{4}R^{ab}\bar{\psi}\gamma ^{cd}\chi \epsilon _{abcd}+\frac{1}{8\ell }%
\mathbb{F}^{ab}\bar{\psi}\gamma ^{cd}\psi \epsilon _{abcd}\right.  \notag \\
&&\left. +\frac{1}{4\ell ^{2}}\mathbb{F}^{ab}\bar{\psi}\gamma ^{cd}\chi
\epsilon _{abcd}+\frac{1}{8\ell }R^{ab}\bar{\chi}\gamma ^{cd}\chi \epsilon
_{abcd}+\frac{1}{8\ell ^{3}}\mathbb{F}^{ab}\bar{\chi}\gamma ^{cd}\chi
\epsilon _{abcd}\right) \,.
\end{eqnarray}%
Thus, let us consider the following full Lagrangian{:}
\begin{eqnarray}
\mathcal{L}_{\text{full}} &=&\mathcal{L}_{\text{bulk}}+\mathcal{L}_{\text{bdy%
}}  \notag \\
&=&\epsilon _{abcd}R^{ab}V^{c}V^{d}+\frac{1}{\ell ^{2}}\epsilon _{abcd}%
\mathbb{F}^{ab}V^{c}V^{d}+4\bar{\psi}\gamma _{a}\gamma _{5}\rho V^{a}  \notag
\\
&&+\frac{4}{\ell }\bar{\psi}\gamma _{a}\gamma _{5}\Phi V^{a}+\frac{4}{\ell
^{2}}\bar{\chi}\gamma _{a}\gamma _{5}\Phi V^{a}+\frac{4}{\ell }\bar{\chi}%
\gamma _{a}\gamma _{5}\rho V^{a}  \notag \\
&&+\frac{1}{2\ell ^{2}}\epsilon _{abcd}V^{a}V^{b}V^{c}V^{d}+\frac{1}{\ell }%
\epsilon _{abcd}\bar{\psi}\gamma ^{ab}\psi V^{c}V^{d}+\frac{2}{\ell ^{2}}%
\epsilon _{abcd}\bar{\chi}\gamma ^{ab}\psi V^{c}V^{d}  \notag \\
&&+\frac{1}{\ell ^{3}}\epsilon _{abcd}\bar{\chi}\gamma ^{ab}\chi
V^{c}V^{d}+\mu \epsilon _{abcd}\left( 2R^{ab}\mathbb{F}^{cd}+\frac{1}{\ell
^{2}}\mathbb{F}^{ab}\mathbb{F}^{cd}\right)  \notag \\
&&+\nu \left( 2\bar{\rho}\gamma _{5}\Phi +\frac{1}{\ell }\bar{\Phi}\gamma
_{5}\Phi +\frac{1}{4}R^{ab}\bar{\psi}\gamma ^{cd}\chi \epsilon _{abcd}+\frac{%
1}{8\ell }\mathbb{F}^{ab}\bar{\psi}\gamma ^{cd}\psi \epsilon _{abcd}\right.
\notag \\
&&\left. +\frac{1}{4\ell ^{2}}\mathbb{F}^{ab}\bar{\psi}\gamma ^{cd}\chi
\epsilon _{abcd}+\frac{1}{8\ell }R^{ab}\bar{\chi}\gamma ^{cd}\chi \epsilon
_{abcd}+\frac{1}{8\ell ^{3}}\mathbb{F}^{ab}\bar{\chi}\gamma ^{cd}\chi
\epsilon _{abcd}\right) \,.  \label{FullL}
\end{eqnarray}%
Now we have to verify the supersymmetry invariance of the full Lagrangian.
{Clearly, $\imath _{\epsilon }d\mathcal{L}_{\text{full}}=0$, since the
boundary terms are total derivatives. Then, considering the condition
\begin{equation}
\delta _{\epsilon }\mathcal{L}_{\text{full}}=\imath _{\epsilon }d\mathcal{L}%
_{\text{full}}+d\left( \imath _{\epsilon }\mathcal{L}_{\text{full}}\right)
=0\,,
\end{equation}
we have just to verify that $\imath _{\epsilon }(\mathcal{L}_{\text{full}})|_{\partial
\mathcal{M}}=0$ is satisfied on the boundary.} Considering \eqref{FullL}, we have%
\begin{eqnarray}
\imath _{\epsilon }\left( \mathcal{L}_{\text{full}}\right) &=& {\epsilon
_{abcd}\imath _{\epsilon }\left( R^{ab}+\frac{1}{\ell ^{2}}
\mathbb{F}^{ab}\right)} V^{c}V^{d}+4\bar{\epsilon}V^{a}\gamma _{a}\gamma
_{5}\rho +4\bar{\psi}V^{a}\gamma _{a}\gamma _{5}\imath _{\epsilon }(\rho )
\notag \\
&&+\frac{4}{\ell }\bar{\epsilon}V^{a}\gamma _{a}\gamma _{5}\Phi +\frac{4}{%
\ell }\bar{\psi}V^{a}\gamma _{a}\gamma _{5}\imath _{\epsilon }(\Phi )+\frac{4%
}{\ell ^{2}}\bar{\chi}V^{a}\gamma _{a}\gamma _{5}\imath _{\epsilon }(\Phi )
\notag \\
&&+\frac{4}{\ell }\bar{\chi}V^{a}\gamma _{a}\gamma _{5}\imath _{\epsilon
}(\rho )+\left( \frac{2}{\ell }\bar{\epsilon}\gamma ^{ab}\psi V^{c}V^{d}+%
\frac{2}{\ell ^{2}}\epsilon _{abcd}\bar{\chi}\gamma ^{ab}\epsilon
V^{c}V^{d}\right) \epsilon _{abcd}  \notag \\
&&+\mu \epsilon _{abcd}\imath _{\epsilon }\left( 2R^{ab}+\frac{1}{\ell ^{2}}%
\mathbb{F}^{ab}\right) \mathbb{F}^{cd}+\mu \epsilon _{abcd}\left( 2R^{ab}+%
\frac{1}{\ell ^{2}}\mathbb{F}^{ab}\right) \imath _{\epsilon }(\mathbb{F}%
^{cd})  \notag \\
&&+2\nu \left[ \imath _{\epsilon }(\bar{\rho})\gamma _{5}\Phi +\bar{\rho}%
\gamma _{5}\imath _{\epsilon }(\Phi )+\frac{1}{\ell }\imath _{\epsilon }(%
\bar{\Phi})\gamma _{5}\Phi \right]  \notag \\
&&+\nu {\epsilon_{abcd}} \left[ \frac{1}{4}\imath _{\epsilon }\left( R^{ab}+\frac{1}{\ell ^{2}}%
\mathbb{F}^{ab}\right) \bar{\psi}\gamma ^{cd}\chi +\frac{1}{4%
}\left( R^{ab}+\frac{1}{\ell ^{2}}\mathbb{F}^{ab}\right) \bar{\epsilon}%
\gamma ^{cd}\chi \right.  \notag \\
&&\left. +\frac{1}{8\ell }\imath _{\epsilon }\left( R^{ab}+\frac{1}{\ell ^{2}%
}\mathbb{F}^{ab}\right) \bar{\chi}\gamma ^{cd}\chi \,+\frac{1%
}{8\ell }\imath _{\epsilon }(\mathbb{F}^{ab})\bar{\psi}\gamma ^{cd}\psi
+\frac{1}{4\ell }\mathbb{F}^{ab}\bar{\epsilon}\gamma
^{cd}\psi \right] \, {.}
\end{eqnarray}%
Then, one can prove that $\left. \frac{\delta {\mathcal{L}}_{\text{full}}}{\delta \mu
^{A}}\right\vert _{\partial \mathcal{M}}=0$ leads to the following
constraints on the boundary{:}
\begin{equation}
{\left\{ \begin{aligned} R^{ab} \vert_{\partial \mathcal{M}} = & \,
-\frac{\nu }{16\mu \ell }\left(\bar{\psi}\gamma
^{ab}\psi \right)_{\partial \mathcal{M}} \,, \\ \mathbb{F}^{ab} \vert_{\partial \mathcal{M}} = & -\frac{1}{2\mu }\left( V^{a}V^{b}+%
\frac{\nu }{4}\bar{\chi}\gamma ^{ab}\psi +\frac{\nu }{8\ell }\bar{\chi}%
\gamma ^{ab}\chi \right)_{\partial \mathcal{M}} \,, \\ \rho \vert_{\partial \mathcal{M}} = & \,
0 \,, \\ \Phi \vert_{\partial \mathcal{M}} = & -\frac{2}{\nu }\left( V^{a}\gamma _{a}\psi +%
\frac{1}{\ell }V^{a}\gamma _{a}\chi \right)_{\partial \mathcal{M}} \,. \end{aligned}\right. }
\label{eombdy}
\end{equation}
{Again, the super curvatures on the boundary result to be fixed to constant values in an enlarged anholonomic basis (given by the supervielbein together with the $1$-form field $\chi$).
Nevertheless, as we have explicitly shown in \eqref{rhparell}, the rheonomic parametrization of these super curvatures results to be well defined in ordinary superspace.}

Then, upon use of \eqref{eombdy} we get
\begin{equation}
\imath _{\epsilon }\left( \mathcal{L}_{\text{full}}\right) |_{\partial
\mathcal{M}}=0\text{ \ }\Leftrightarrow \quad \frac{\nu }{8\mu }+\frac{4}{%
\nu }=2\,.
\end{equation}%
This condition can be written as
\begin{equation}
\nu =8\mu \left( 1+h\right) ,\qquad h^{2}=1-\frac{1}{2\mu };\qquad \left(
\nu \neq 0\Rightarrow h\neq -1\right) \,.  \label{solufinal}
\end{equation}%
One can see that setting $h=0$, we obtain
\begin{equation}
\mu =\frac{1}{2}\quad \Rightarrow \quad \nu =4\,.  \label{adsLvalues}
\end{equation}%
Remarkably, with these values for $\mu $ and $\nu $, the full Lagrangian %
\eqref{FullL} can be written à la MacDowell-Mansouri as
\begin{equation}
\mathcal{L}_{\text{full}}=\mathfrak{R}^{ab}\mathfrak{F}^{cd}\epsilon _{abcd}+%
\frac{1}{2\ell ^{2}}\mathfrak{F}^{ab}\mathfrak{F}^{cd}\epsilon _{abcd}+8\bar{%
\Omega}\gamma _{5}\rho +\frac{4}{\ell }\bar{\Omega}\gamma _{5}\Omega \,\,,  \label{finalFULLLagr}
\end{equation}%
written exactly in terms of the curvatures (\ref{sc1}).

We have thus shown that the supersymmetric boundary terms given in
\eqref{bdyLagrangian} allow to recover the supersymmetry invariance of our
supergravity model in the presence of a non-trivial boundary. One can notice{, using \eqref{eombdy},} that the super curvatures (\ref{sc1}) vanish at the boundary.

The flat limit $\ell \rightarrow \infty $ is now well defined in the
MacDowell-Mansouri formalism. Indeed, unlike the {case of the full Lagrangian obtained for} $\mathfrak{osp}\left(4|1\right) $ supergravity {in the presence of a non-trivial boundary \cite{AD}}, the vanishing cosmological constant limit of the
full Lagrangian {\eqref{finalFULLLagr}} reproduces appropriately the flat supergravity model {with boundary} that we
have {considered} initially. In particular, in this case not only the bulk Lagrangians are well related
through the flat limit but also the boundary contributions. Furthermore, {one can see that} the super-Maxwell curvatures {(\ref{FabMcurv}), (\ref{ChiMcurv}), and (\ref{Mcurv}) are recovered} as a flat limit of
the super curvatures (\ref{sc1}). It is then natural to expect that the
minimal Maxwell superalgebra (\ref{sM}) appears as a flat limit of a
deformation of the $\mathfrak{osp}\left( 4|1\right) $ superalgebra.

\subsubsection{AdS-Lorentz supersymmetry}

The (anti)commutation relations of {the superalgebra related to}
the super curvatures (\ref{sc1}) are given by
\begin{eqnarray}
\left[ J_{ab},J_{cd}\right] &=&\eta _{bc}J_{ad}-\eta _{ac}J_{bd}-\eta
_{bd}J_{ac}+\eta _{ad}J_{bc}\,,  \notag \\
\left[ J_{ab},Z_{cd}\right] &=&\eta _{bc}Z_{ad}-\eta _{ac}Z_{bd}-\eta
_{bd}Z_{ac}+\eta _{ad}Z_{bc}\,,\medskip  \notag \\
\left[ Z_{ab},Z_{cd}\right] &=&\dfrac{1}{\ell ^{2}}\left( \eta
_{bc}Z_{ad}-\eta _{ac}Z_{bd}-\eta _{bd}Z_{ac}+\eta _{ad}Z_{bc}\right) \,,
\label{AdSL} \\
\left[ J_{ab},P_{c}\right] &=&\eta _{bc}P_{a}-\eta _{ac}P_{b}\,,\qquad \left[
Z_{ab},P_{c}\right] =\dfrac{1}{\ell ^{2}}\left( \medskip \eta
_{bc}P_{a}-\eta _{ac}P_{b}\right) \,,  \notag \\
\left[ P_{a},P_{b}\right] &=&Z_{ab}\,,  \notag
\end{eqnarray}%
\begin{eqnarray}
\left[ J_{ab},Q_{\alpha }\right] &=&-\dfrac{1}{2}\left( \gamma _{ab}Q\right)
_{\alpha }\,,\qquad \left[ J_{ab},\Sigma _{\alpha }\right] =-\frac{1}{2}%
\left( \gamma _{ab}\Sigma \right) _{\alpha }\,,  \notag \\
\left[ Z_{ab},Q_{\alpha }\right] &=&-\dfrac{1}{2\ell }\left( \gamma
_{ab}\Sigma \right) _{\alpha }\,,\qquad \left[ Z_{ab},\Sigma _{\alpha }%
\right] =-\frac{1}{2\ell ^{2}}\left( \gamma _{ab}\Sigma \right) _{\alpha }\,,
\notag \\
\left[ P_{a},Q_{\alpha }\right] &=&-\dfrac{1}{2}\left( \gamma _{a}\Sigma
\right) _{\alpha }\,,\qquad \left[ P_{a},\Sigma _{\alpha }\right] =-\dfrac{1%
}{2\ell }\left( \gamma _{a}\Sigma \right) _{\alpha }\,,  \notag \\
\left\{ Q_{\alpha },Q_{\beta }\right\} &=&-\frac{1}{2\ell }\left( \gamma
^{ab}C\right) _{\alpha \beta }J_{ab}+\left( \gamma ^{a}C\right) _{\alpha
\beta }P_{a}\,, \label{SUSYpart} \\
\left\{ Q_{\alpha },\Sigma _{\beta }\right\} &=&-\frac{1}{2}\left( \gamma
^{ab}C\right) _{\alpha \beta }Z_{ab}+\frac{1}{\ell }\left( \gamma
^{a}C\right) _{\alpha \beta }P_{a}\,,  \notag \\
\left\{ \Sigma _{\alpha },\Sigma _{\beta }\right\} &=&-\frac{1}{2\ell }%
\left( \gamma ^{ab}C\right) _{\alpha \beta }Z_{ab}+\frac{1}{\ell ^{2}}\left(
\gamma ^{a}C\right) _{\alpha \beta }P_{a}\,,  \notag
\end{eqnarray}%
with $a=0,1,2,3$, $\alpha =1,2,3,4$, and where the generators $J_{ab}$, $%
P_{a}$, $Z_{ab}$, $Q$ and $\Sigma $ are respectively dual to the $1$-form
fields $\omega ^{ab}$, $V^{a}$, $A^{ab}$, $\psi $, and $\chi $.
Interestingly, the present superalgebra corresponds to an alternative
supersymmetric extension of the AdS-Lorentz symmetry (given by eq. (\ref%
{AdSL})). Such bosonic algebra, also known as a semisimple extension of the
Poincaré symmetry, was first presented in \cite{SS, GKL}. Then, it was generalized to a family of $\mathfrak{C}_{k}$ algebras \cite{CDMR,
Durka} which have been useful to recover diverse (pure) Lovelock theories
from CS and BI gravity theories \cite{CDIMR, CMR, CR3}.

Although it contains two spinorial charges as the minimal AdS-Lorentz superalgebras introduced in \cite{CRS}, {the superalgebra given by eqs. \eqref{AdSL} and \eqref{SUSYpart}} does not require additional
bosonic generators {with respect to the ones of the AdS-Lorentz algebra} in order to satisfy the Jacobi identities. The closure of
this superalgebra is guaranteed by the explicit form of the anticommutators {in \eqref{SUSYpart}.}
Indeed, any subtle deformation of {the aforementioned} anticommutators requires to introduce
additional bosonic generators as in \cite{BGKL2, CRS, BR}. On the other
hand, it is important to signalize that the supersymmetrization of the
AdS-Lorentz algebra is not unique. In particular, a {super AdS-Lorentz}
symmetry with one spinor charge has also been considered in \cite{SS, SS2,
FISV}.

Note that in the limit $\ell \rightarrow \infty $ the above superalgebra gives exactly the super-Maxwell algebra (\ref{sM}), without any auxiliary generator.

As a final remark, it is important to point out that in \cite{CIRR} a
supergravity model in presence of a non-trivial boundary for the AdS-Lorentz {superalgebra} with one spinor charge has been presented. Nevertheless, such construction
does not allow for a proper flat limit. In particular, although it is possible to recover a non-standard Maxwell superalgebra from the super AdS-Lorentz one considered in \cite{CIRR}, the exotic anticommutation relation of the
non-standard Maxwell does not allow a proper construction of a supergravity
action. Such problematic comes from the fact that the {four-momentum
generators no longer appear as a result of the anticommutator of the fermionic generators}. It seems that, as we have shown here,
the only possibility to have a well defined vanishing cosmological constant
limit for the AdS-Lorentz supergravity is to consider an additional
fermionic charge. Remarkably, such flat limit allows to recover the usual
flat supergravity with non-trivial boundary contributions.

\section{Discussion}

{In this paper, we have studied} the supersymmetry invariance of flat supergravity in presence of a non-trivial boundary. We {have shown} that supersymmetry invariance is achieved by adding proper topological terms in which additional gauge fields are considered. We {have found} that the full Lagrangian can be rewritten à la MacDowell-Mansouri in terms of Maxwell super curvatures. Interestingly, the bulk Lagrangian corresponds to the usual flat supergravity Lagrangian. Although the extra fields only appear {in} the boundary {contributions,} they are essential to recover supersymmetry invariance of the full Lagrangian. Furthermore, as the topological term in pure gravity allows to regularize the action, one could argue that the presence of the new gauge fields in the boundary would allow to regularize the supergravity action in the holographic renormalization language.

We have also explored the possibility of introducing a well-defined vanishing cosmological constant limit in a supergravity model in order to recover the Lagrangian obtained in the case of flat spacetime with a non-trivial boundary. As we have discussed, flat supergravity with boundary cannot be naively obtained from $\mathfrak{osp}(4|1)$ supergravity with boundary (MacDowell{-}Mansouri action). Indeed, this require to consider an enlarged supergravity with a generalized cosmological constant which has been obtained using the rheonomic approach. We {have shown} that, as in the flat case, supersymmetry invariance requires to add appropriate topological terms. The full supergravity obtained {within this procedure} can be rewritten in terms of the super curvatures of a particular AdS-Lorentz superalgebra. Remarkably, the flat limit of the full Lagrangian properly reproduces flat supergravity on a manifold with boundary.

{To our knowledge, this is the first report showing that the supersymmetry invariance of a supergravity theory on a manifold with boundary is restored by adding new gauge fields in the boundary. However, the introduction of new gauge fields which enlarge the symmetry group has been already considered in the literature with different purposes. Of particular interest are the Maxwell and AdS-Lorentz algebras. Although such symmetries have been particularly useful to relate diverse gravity theories \cite{EHTZ, GRCS, CPRS1, CPRS2, CPRS3, CDIMR, CMR, CR3}, the dynamical implication of considering an enlarged set of fields has been poorly explored. In early works, in order to interpret the Maxwell (super)algebra, and the corresponding (super)group, a Maxwell-invariant particle model on the extended spacetime was studied \cite{GKL, BGKL2, Bonanos:2008kr, Bonanos:2008ez, Gomis:2009vm, Gibbons:2009me}. In particular, it was shown that the interaction term described by a Maxwell-invariant $1$-form introduces new tensor degrees of freedom that, in the equations of motion, play the role of a background electromagnetic field which is constant on-shell and leads to a closed, Maxwell-invariant $2$-form. On the other hand, in \cite{AKL} the Maxwell algebra was exploited in order introduce the cosmological term in Einstein gravity in an alternative way. This appeared in a generalized form, with a dependence on the additional gauge fields $A^{ab}_{\mu}$ associated with the generators $Z_{ab}$, giving an interpretation of the Maxwell algebra in the context of inflation, since the latter can also be driven by suitably coupled vector fields. Such feature was then extended to supergravity in \cite{CRS, CIRR, PR2}, offering, in particular, an alternative way of introducing a generalized cosmological term in this context.}

{Moreover, recently it was shown in three spacetime dimensions that the presence of the extra fields in the Chern-Simons action invariant under the Maxwell algebra influences the vacuum energy and the vacuum angular momentum of the stationary configuration \cite{CMMRSV}. In addition, it has been shown that the boundary dynamics of such gravity theory is described by an enlargement and deformation of the $\mathfrak{bms}_3$ algebra with three central charges. On the other hand, considering suitable boundary conditions, a semi-simple enlargement of the $\mathfrak{bms}_3$ algebra appears as the asymptotic symmetry at null infinity of a Chern-Simons gravity action based on the AdS-Lorentz algebra \cite{CMRSV}. The extension of these results to supergravity remains unknown and it would be interesting to generalize the study in four dimensions.}

{Furthermore, as we have already mentioned, the presence of an extra (that is besides $Q$) fermionic generator (we called it $\Sigma$ in the present work) dual to a spinor $1$-form (here, $\chi$) is naturally involved in the minimal supersymmetrization of the Maxwell algebra.
On the other hand, other relevant superalgebras containing two fermionic charges were introduced and deeply analyzed in \cite{Ravera, DAuria:1982uck, Green:1989nn, Andrianopoli:2016osu, Andrianopoli:2017itj, PR}: Precisely, the D'Auria-Fr\'{e} superalgebra, introduced in \cite{DAuria:1982uck} and subsequently further analyzed in \cite{Andrianopoli:2016osu, Andrianopoli:2017itj}, underlying the Free Differential Algebra of $D=11$ supergravity, the Green algebra \cite{Green:1989nn}, in the context of superstring, and Maxwell-type superalgebras recently shown to be related to $D = 4$ and $D = 11$ supergravity \cite{Ravera, PR}. Then, we conjecture the existence of possible relations among the new $1$-form fields $A^{ab}$ and $\chi$ appearing in our model (and, eventually, in extensions to higher dimensions) and the extra fields appearing in the aforementioned theories. This could shed some light on the physical and group theoretical role played by $A^{ab}$ and $\chi$.
Concerning supergravity theories, an interesting relation has already been found in \cite{Ravera} (following previous studies presented for the first time in \cite{DAuria:1982uck} and further analyzed in \cite{PR, Andrianopoli:2016osu, Andrianopoli:2017itj}), where it was shown that the Maxwell superalgebra in four dimensions, whose corresponding super curvatures are the ones given in \eqref{FabMcurv}, \eqref{ChiMcurv}, and \eqref{Mcurv}, can be interpreted as a hidden superalgebra underlying $\mathcal{N}=1$, $D=4$ supergravity extended to include a $2$-form gauge potential associated to a $2$-index antisymmetric tensor (in this scenario, the theory is appropriately discussed in
the context of Free Differential Algebras, an extension of the Maurer-Cartan equations to involve higher-degree differential forms). The same extra spinor dual to the nilpotent fermionic generator whose presence is crucial for writing a supersymmetric extension of the Maxwell algebra turned out to be fundamental ingredients also to reproduce the $D = 4$ Free Differential Algebra on ordinary superspace. A future work on this side would consist in extending the analysis to higher dimensions, clarifying, in this context, the physical (and topological) role of the field $A^{ab}$, dual to the almost-central (in the sense that it commutes with all the superalgebra but the Lorentz generators), bosonic generator $Z_{ab}$ that could be understood as a $2$-brane charge (probably under a generalized perspective), source of a $3$-form gauge potential. Furthermore, the study could be extended to the case in which an explicit scale is switched on.}

It is important to observe that, as in \cite{AD, CIRR, BR}, the full supergravity Lagrangian can be rewritten à la MacDowell-Mansouri. This would suggest, as {it} was pointed out in the bosonic case \cite{MO, MOT}, a superconformal structure which is an interesting additional motivation {to our study}. On the other hand, our particular approach in which we consider additional gauge {fields} in the boundary could be useful in order to explore the supersymmetry invariance of supergravities with boundary coupled to matter or in higher dimensions. Indeed, it would be interesting to explore the boundary contributions required to recover supersymmetry invariance of a general matter coupled $\mathcal{N}=2$ supergravity in four dimensions \cite{ABCDFFM, ACDRT}. Naturally, one could consider first a supergravity theory coupled to scalar field. {The extension} of our results to $\mathcal{N}=2$ is more subtle and will be studied in a future work.

{Finally, let us highlight that in the present work we have shown that the supersymmetry invariance of a supergravity theory on a manifold with boundary is restored by adding new gauge fields in the boundary, concentrating just on the four-dimensional theory to show that the supersymmetry of the four-dimensional Lagrangian given by bulk plus boundary contributions can indeed be recovered. In this context, the supersymmetry invariance of the $D=4$ theory is achieved without requiring Dirichlet boundary conditions on the fields at the boundary; rather we have found that the boundary values of the super curvatures are dynamically fixed to constant values in an enlarged anholonomic basis (indeed, on the boundary they are fixed in terms of not only the supervielbein, but also of the extra $1$-form field $\chi$). Nevertheless, as we have explicitly shown, the rheonomic parametrization of these super curvatures results to be well defined in ordinary superspace.}

{However, here we shall mention that what commonly happens is that the insertion of a spacetime boundary breaks a part of the boundary supersymmetries, and thus one would expect that our boundary conditions break a part of the bulk supersymmetries on the boundary. The analysis of the symmetry breaking pattern in this context would require the study of the theory living on the boundary and, in particular, the explicit three-dimensional description of the equations (precisely, of the boundary conditions on the super field-strength) we have obtained in $D=4$.
Thus, one could investigate on the boundary theory produced in our framework: In order to discuss the theory living on the boundary (in the spirit of the holographic correspondence) one should set the boundary at $r \rightarrow \infty$, where $r$ is the radial coordinate, and study the asymptotic limit $r \rightarrow \infty$ of the $D=3$ equations on the boundary (in particular, one should properly choose the boundary behavior of the $D=4$ fields which relates them to the $D = 3$ ones and perform the asymptotic limit $r \rightarrow \infty$). The explicit three-dimensional description of the equations we have obtained in $D=4$ would depend on the general symmetry properties of the theory on the boundary.}

{In the scenario of our work, where we have the presence of a non-trivial boundary of spacetime (meaning that the boundary is not thought as set at infinity and thus the fields do not asymptotically vanish) and of extra bosonic and fermionic $1$-form fields ($A^{ab}$ and $\chi$, respectively), we conjecture that the related boundary theory (at least, concerning the enlarged supergravity model involving the scale parameter $\ell$; then, one could further study the limit $\ell \rightarrow \infty$ in this context) could feature some generalization of deformed locally AdS$_3$ geometries, due to the presence of extra $D=4$ fields from the very beginning.}

{Moreover, we conjecture that such an analysis could also shed further light on the study done in \cite{ACDT}, where the authors found unexpected intriguing
relations between $\mathcal{N}=2$, $D=4$ supergravity and a three-dimensional theory describing the properties of graphene and featuring ``unconventional local supersymmetry''.
Besides, it would be interesting to further discuss our results in light of the ones obtained in \cite{Ciambelli:2018wre} in the context of flat holography and Carrollian fluids, where the authors demonstrated, using the derivative expansion of fluid/gravity correspondence, that a holographic description of four-dimensional asymptotically locally flat spacetimes is reached smoothly from the zero cosmological constant limit of AdS holography. Intriguingly, from the boundary perspective the vanishing of the bulk cosmological constant appears as the zero velocity of light limit, setting how Carrollian geometry
emerges in flat holography.
The analysis of the boundary theory produced in our framework, together with a deeper investigation on the role played by the spinor $1$-form field $\chi$ and the bosonic $1$-form field $A^{ab}$ appearing in our model in this context, will be will the subject of a future work.}

\section{Acknowledgment}

The authors wish to thank L. Andrianopoli, R. D'Auria and M. Trigiante for
enlightening discussions and their kind hospitality at DISAT of Politecnico
di Torino (Italy), where the main discussion of this work was done. This work
was supported by the Chilean FONDECYT Projects N$^{\circ }$3170437 (P.C.)
and N$^{\circ }$3170438 (E.R.).


\begin{thebibliography}{99}
\bibitem{York} J.W. York, Jr., \textit{Role of conformal three geometry in
the dynamics of gravitation}, Phys. Rev. Lett. \textbf{28} (1972) 1082.

\bibitem{GH} G.W. Gibbons, S.W. Hawking, \textit{Action Integrals and
Partition Functions in Quantum Gravity}, Phys. Rev. D \textbf{15} (1977)
2752.

\bibitem{BY} J.D. Brown, J.W. York, Jr., \textit{Quasilocal energy and
conserved charges derived from the gravitational action}, Phys. Rev. D%
\textbf{47 }(1993) 1407. [gr-qc/9209012].

\bibitem{HW} P. Horava, E. Witten, \textit{Eleven-dimensional supergravity
on a manifold with boundary}, Nucl. Phys. B \textbf{475} (1996) 94.
[hep-th/9603142].

\bibitem{Maldacena} J.M. Maldacena, \textit{The Large N limit of
superconformal field theories and supergravity}, Int. J. Theor. Phys.
\textbf{38} (1999) 1113. [hep-th/9711200].

\bibitem{GKP} S.S. Gubser, I.R. Klebanov, A.M. Polyakov, \textit{Gauge
theory correlators from noncritical string theory}, Phys. Lett. B \textbf{428%
} (1998) 105. [hep-th/9802109].

\bibitem{Witten} E. Witten, \textit{Anti-de Sitter space and holography},
Adv. Theor. Math. Phys. \textbf{2} (1998) 253. [hep-th/9802150].

\bibitem{AGMOO} O. Aharony, S.S. Gubser, J.M. Maldacena, H. Ooguri, Y. Oz,
\textit{Large-N field theories, string theory and gravity}, Phys. Rept.
\textbf{323} (2000) 183. [hep-th/990511].

\bibitem{HF} E. D'Hoker, D.Z. Freedman, \textit{Supersymmetric gauge
theories and the AdS/CFT correspondence}. [hep-th/0201253].

\bibitem{CM} {G. Compère and D. Marolf, \textit{Setting the boundary free in AdS/CFT}, Class.\ Quant.\ Grav.\  {\bf 25} (2008) 195014. arXiv:0805.1902 [hep-th].}

\bibitem{AC} {A. J. Amsel and G. Compère, \textit{Supergravity at the boundary of AdS supergravity}, Phys.\ Rev.\ D {\bf 79} (2009) 085006. arXiv:0901.3609 [hep-th].}

\bibitem{BK} V. Balasubramanian, P. Kraus, \textit{A stress tensor for
Anti-de Sitter gravity}, Commun. Math. Phys. \textbf{208} (1999) 413.
[hep-th/9902121].

\bibitem{BVV} J. de Boer, E.P. Verlinde, H.L. Verlinde, \textit{On the
holographic renormalization group}, JHEP \textbf{0008} (2000) 003.
[hep-th/9912012].

\bibitem{VV} E.P. Verlinde, H.L. Verlinde, \textit{RG flow, gravity and the
cosmological constant}, JHEP \textbf{0005} (2000) 034. [hep-th/9912018].

\bibitem{Boer} J. de Boer, \textit{The Holographic renormalization group},
Fortsch. Phys. \textbf{49} (2001) 339. [hep-th/0101026].

\bibitem{HSS} S. de Haro, S.N. Soludkhin, K. Skenderis, \textit{Holographic
reconstruction of space-time and renormalization in the AdS/CFT
correspondence}, Commun. Math. Phys. \textbf{217} (2001) 595.
[hep-th/0002230].

\bibitem{Skenderis} K. Skenderis, \textit{Lecture notes on holographic
renormalization}, Class. Quant. Grav. \textbf{19} (2002) 5849
[hep-th/0209067].

\bibitem{LS} A. Lawrence, A. Sever, \textit{Holography and renormalization in Lorentzian signature}, JHEP \textbf{0610} (2006) 013. [hep-th/0606022].

\bibitem{HP} I. Heemskerk, J. Polchinski, \textit{Holographic and Wilsonian Renormalization Groups}, JHEP \textbf{1106} (2011) 031. arXiv:1010.1264 [hep-th].

\bibitem{ABD} D. Astefanesei, N. Banerjee, S. Dutta, \textit{(Un)attractor black holes in higher derivative AdS gravity}, JHEP \textbf{0811} (2008) 070. arXiv:0806.1334 [hep-th].

\bibitem{BRadu} Y. Brihaye, E. Radu, \textit{Black objects in the Einstein-Gauss-Bonnet theory with constant and the boundary counterterm method}, JHEP \textbf{0809} (2008) 006. arXiv:0806.1396 [gr-qc].

\bibitem{AACM} A. Anabalón, D. Astefanesei, D. Choque, C. Martínez, \textit{Trace anomaly and Counterterms in Designer Gravity}, JHEP \textbf{03} (2016) 117. arXiv:1511.08759 [hep-th].

\bibitem{ABCR} D. Astefanesei, R. Ballesteros, D. Choque, R. Rojas, \textit{Scalar charges and the first law of black hole thermodynamics}, Phys. Lett. B \textbf{782} (2018) 47. arXiv:1803.11317 [hep-th].

\bibitem{EKK} G. Esposito, A.Y. Kamenshchik, K. Kirsten, \textit{One loop
effective action for Euclidean Maxwell theory on manifolds with boundary},
Phys. Rev. D \textbf{54} (1996) 7328 [hep-th/9606132].

\bibitem{Moss} I.G. Moss, \textit{Boundary terms for eleven-dimensional
supergravity and M theory}, Phys. Lett. B \textbf{577} (2003) 71.
[hep-th/0308159].

\bibitem{NV} P. van Nieuwenhuizen, D.V. Vassilevich, \textit{Consistent
boundary conditions for supergravity}, Class. Quant. Grav. \textbf{22}
(2005) 5029. [hep-th/0507172].

\bibitem{B} D.V. Belyaev, \textit{Boundary conditions in supergravity on a
manifold with boundary}, JHEP \textbf{0601} (2006) 047. [hep-th/0509172].

\bibitem{NRVW} {P. van Nieuwenhuizen, A. Rebhan, D. V. Vassilevich and R. Wimmer, \textit{Boundary terms in supergravity and supersymmetry}, Int.\ J.\ Mod.\ Phys.\ D {\bf 15} (2006) 1643. [hep-th/0606075].}

\bibitem{BN} {D. V. Belyaev and P. van Nieuwenhuizen, \textit{Simple d=4 supergravity with a boundary}, JHEP {\bf 0809} (2008) 069. arXiv:0806.4723 [hep-th].}

\bibitem{BN1} D.V. Belyaev, P. van Nieuwenhuizen, \textit{Tensor calculus
for supergravity on a manifold with boundary}, JHEP \textbf{0802} (2008)
047. arXiv:0711.2272 [hep-th].

\bibitem{GN} D. Grumiller, P. van Nieuwenhuizen, \textit{Holographic
counterterms from local supersymmetry without boundary conditions}, Phys.
Lett. B \textbf{682} (2010) 462. arXiv:0908.3486 [hep-th].

\bibitem{HPSS} P.S. Howe, T.G. Pugh, K.S. Stelle, C. Strickland-Constable,
\textit{Ectoplasm with an Edge}, JHEP \textbf{1108} (2011) 081.
arXiv:1104.4387 [hep-th].

\bibitem{AD} L. Andrianopoli, R. D'Auria, \textit{N=1 and N=2
pure supergravities on a manifold with boundary}, JHEP \textbf{1408} (2014)
012. arXiv:1405.2010 [hep-th].

\bibitem{PKS} L. Di Pietro, N. Klinghoffer, I. Shamir, \textit{On
Supersymmetry, Boundary Actions and Brane Charges}, JHEP \textbf{1602 }%
(2016) 163. arXiv:1502.05976 [hep-th].

\bibitem{CIRR} P.K. Concha, M.C. Ipinza, L. Ravera, E.K. Rodríguez, \textit{%
On the supersymmetric extension of Gauss-Bonnet like gravity}, JHEP \textbf{%
09} (2016) 007. arXiv:1607.00373 [hep-th].

\bibitem{FPPW} D.Z. Freedman, K. Pilch, S.S. Pufu, N.P. Warner, \textit{%
Boundary Terms and Three-Point Functions: An AdS/CFT Puzzle Resolved}, JHEP
\textbf{1706} (2017) 053. arXiv:1611.01888 [hep-th].

\bibitem{BCMS} P. Benetti Genolini, D. Cassani, D. Martelli, J. Sparks,
\textit{Holographic renormalization and supersymmetry}, JHEP \textbf{1702}
(2017) 132. arXiv:1612.06761 [hep-th].

\bibitem{AAGT} A. Anabalón, D. Astefanesei, A. Gallerati, M. Trigiante, \textit{Hairy Black Holes and Duality in an Extended Supergravity Model}, JHEP \textbf{04} (2018) 058. arXiv:1712.06971 [hep-th].

\bibitem{ACDT} L. Andrianopoli, B.L. Cerchiai, R. D'Auria, M. Trigiante,
\textit{Unconventional supersymmetry at the boundary of }$AdS_{4}\mathit{\ }$%
\textit{supergravity}, JHEP \textbf{1804} (2018) 007. arXiv:1801.08081
[hep-th].

\bibitem{BR} A. Baunadi, L. Ravera, \textit{Generalized AdS-Lorentz deformed
supergravity on a manifold with boundary}, Eur. Phys. J. Plus. \textbf{133} (2018) 514. arXiv:1803.08738 [hep-th].

\bibitem{ACOTZ} R. Aros, M. Contreras, R. Olea, R. Troncoso, J. Zanelli,
\textit{Conserved charges for gravity with locally AdS asymptotics}, Phys.
Rev. Lett. \textbf{84} (2000) 1647. [gr-qc/9909015].

\bibitem{ACOTZ2} R. Aros, M. Contreras, R. Olea, R. Troncoso, J. Zanelli,%
\textit{\ Conserved charges for even dimensional asymptotically AdS gravity
theories}, Phys. Rev. D \textbf{92} (2000) 044002. [hep-th/9912045].

\bibitem{MOTZ} P. Mora, R. Olea, R. Troncoso, J. Zanelli, \textit{Finite
action principle for Chern-Simons AdS gravity}, JHEP \textbf{0406} (2004)
036. [hep-th/0405267].

\bibitem{Olea} R. Olea, \textit{Mass, angular momentum and thermodynamics in
four-dimensional Kerr-AdS black holes}, JHEP \textbf{0506} (2005) 023.
[hep-th/0504233].

\bibitem{JKMO} D.P. Jatkar, G. Kofinas, O. Miskovic, R. Olea, \textit{%
Conformal Mass in AdS gravity}, Phys. Rev. D\textbf{89} (2014) 124010.
arXiv:1404.1411 [hep-th].

\bibitem{MM} S.W. MacDowell, F. Mansouri, \textit{Unified Geometric Theory
of Gravity and Supergravity}, Phys. Rev. Lett. \textbf{38} (1977) 739.

\bibitem{BGKL} S. Bonanos, J. Gomis, K. Kamimura, J. Lukierski, \textit{Maxwell superalgebra and superparticle in constant Gauge background}.Phys.
Rev. Lett. \textbf{104} (2010) 090401. arXiv:0911.5072 [hep-th].

\bibitem{SS} D.V. Soroka, V.A. Soroka,\textit{\ Semi-simple extension of the
(super) Poincaré algebra, }Adv. High Energy Phys. \textbf{2009} (2009)
234147. [hep-th/0605251].

\bibitem{GKL} J. Gomis, K. Kamimura, J. Lukierski, \textit{Deformations of
Maxwell algebra and their Dynamical Realizations}, JHEP \textbf{0908} (2009)
039. arXiv:0906.4464 [hep-th].

\bibitem{CDF} L. Castellani, R. D'Auria, P. Fré, \textit{Supergravity and superstrings: A geometric prespective. Vol. 1 and 2}, World Scientific, Singapore (1991).

\bibitem{CRS} P.K. Concha, E.K. Rodríguez, P. Salgado, \textit{Generalized
supersymmetric cosmological term in N=1 Supergravity}, JHEP \textbf{08}
(2015) 009. arXiv:1504.01898 [hep-th].

\bibitem{BCR} H. Bacry, P. Combe, J.L. Richard, \textit{Group-theoretical
analysis of elementary particles in an external electromagnetic fields. 1.
The relativistic particle in a constant and uniform field}, Nuovo Cim. A
\textbf{67} (1970) 267.

\bibitem{Schrader} R. Schrader, \textit{The Maxwell group and the quantum
theory of particles in classical homogeneous electromagnetic fields},
Fortsch. Phys. \textbf{20} (1972) 701.

\bibitem{EHTZ} J.D. Edelstein, M. Hassaine, R. Troncoso, J. Zanelli, \textit{%
Lie-algebra expansions, Chern-Simons theories and the Einstein-Hilbert
Lagrangian}, Phys. Lett. B \textbf{640} (2006) 278. [hep-th/0605174].

\bibitem{GRCS} F. Izaurieta, E. Rodríguez, P. Minning, P. Salgado, A. Perez,
\textit{Standard General Relativity from Chern-Simons Gravity}, Phys. Lett.
B \textbf{678} (2009) 213. arXiv:0905.2187 [hep-th].

\bibitem{CPRS1} P.K. Concha, D.M. Peñafiel, E.K. Rodríguez, P. Salgado,
\textit{Even-dimensional General Relativity from Born-Infeld gravity}, Phys.
Lett. B \textbf{725} (2013) 419. arXiv:1309.0062 [hep-th].

\bibitem{CPRS2} P.K. Concha, D.M. Peñafiel, E.K. Rodríguez, P. Salgado,
\textit{Chern-Simons and Born-Infeld gravity theories and Maxwell algebras
type}, Eur. Phys. J. C \textbf{74} (2014) 2741. arXiv:1402.0023 [hep-th].

\bibitem{CPRS3} P.K. Concha, D.M. Peñafiel, E.K. Rodríguez, P. Salgado,
\textit{Generalized Poincaré algebras and Lovelock-Cartan gravity theory},
Phys. Lett. B \textbf{742} (2015) 310. arXiv:1405.7078 [hep.th].

\bibitem{SSV} P. Salgado, R.J. Szabo, O. Valdivia, \textit{Topological
gravity and transgression holography}, Phys. Rev. D\textbf{89} (2014)
084077. arXiv:1401.3653 [hep-th].

\bibitem{HR} S. Hoseinzadeh, A. Rezaei-Aghdam, \textit{(2+1)-dimensional
gravity from Maxwell and semisimple extension of the Poincaré gauge
symmetric models}, Phys. Rev. D\textbf{90} (2014) 084008. arXiv:1402.0320
[hep-th].

\bibitem{CCFRS} R. Caroca, P. Concha, O. Fierro, E. Rodríguez, P.
Salgado-Rebolledo, \textit{Generalized Chern-Simons higher-spin gravity
theories in three dimensions}, Nucl. Phys. B \textbf{934} (2018) 240.
arXiv:1712.09975 [hep-th].

\bibitem{AFGHZ} L. Avilés, E. Frodden, J. Gomis, D. Hidalgo, J. Zanelli,
\textit{Non-Relativistic Maxwell Chern-Simons Gravity}, JHEP \textbf{05}
(2018) 047. arXiv:1802.08453 [hep-th].

\bibitem{CMMRSV} P. Concha, N. Merino, O. Miskovic, E. Rodríguez, P.
Salgado-Rebolledo, O. Valdivia, \textit{Extended asymptotic symmetries of
three-dimensional gravity in flat space}, arXiv:1805.08834 [hep-th].

\bibitem{CFRS} P.K. Concha, O. Fierro, E.K. Rodríguez, P. Salgado, \textit{%
Chern-Simons supergravity in D=3 and Maxwell superalgebra}, Phys. Lett. B
\textbf{750} (2015) 117. arXiv:1507.02335 [hep-th].

\bibitem{CFR} P.K. Concha, O. Fierro, E.K. Rodríguez, \textit{Inönü-Wigner
contraction and D=2+1 supergravity}, Eur. Phys. J. C \textbf{77} (2017) 48.
arXiv:1611.05018 [hep-th].

\bibitem{CPR} P. Concha, D.M. Peñafiel, E. Rodríguez, \textit{On the Maxwell
supergravity and flat limit in 2+1 dimensions}, Phys. Lett. B \textbf{785} (2018) 247. arXiv:1807.00194 [hep-th].

\bibitem{AILW} J.A. de Azcarraga, J.M. Izquierdo, J. Lukierski, M.
Woronowicz, \textit{Generalizations of Maxwell (super)algebras by the
expansion method}, Nucl. Phys. B \textbf{869} (2013) 303. arXiv:1210.1117
[hep-th].

\bibitem{AI} J.A. de Azcarraga, J.M. Izquierdo, \textit{Minimal D=4
supergravity from superMaxwell algebra}, Nucl. Phys. B \textbf{885} (2014)
34. arXiv:1403.4128 [hep-th].

\bibitem{CR1} P.K. Concha, E.K. Rodríguez, \textit{Maxwell superalgebras and
Abelian semigroup expansion}, Nucl. Phys. B \textbf{886} (2014) 1128.
arXiv:1405.1334 [hep-th].

\bibitem{CR2} P.K. Concha, E.K. Rodríguez, \textit{N=1 Supergravity and
Maxwell superalgebras}, JHEP \textbf{1409} (2014) 090. arXiv:1407.4635
[hep-th].

\bibitem{PR} D.M. Peñafiel, L. Ravera, \textit{On the Hidden Maxwell
Superalgebra underlying D=4 Supergravity}, Fortsch. Phys. \textbf{65} (2017)
1700005. arXiv:1701.04234 [hep-th].

\bibitem{Ravera} L. Ravera, \textit{Hidden role of Maxwell superalgebras in
the free differential algebras of }$D=4$\textit{\ and }$D=11$\textit{\
supergravity}, Eur. Phys. J. C \textbf{78} (2018) 211. arXiv:1801.08860
[hep-th].

\bibitem{AKL} J.A. de Azcarraga, K. Kamimura, J. Lukierski, \textit{%
Generalized cosmological term from Maxwell symmetries}, Phys. Rev. D\textbf{%
83} (2011) 124036. arXiv:1012.4402 [hep-th].

\bibitem{PR2} D.M. Peñafiel, L. Ravera, \textit{Generalized cosmological
term in D=4 supergravity from a new AdS-Lorentz superalgebra}, Eur. Phys. J. C \textbf{78} (2018) 945.
arXiv:1807.07673 [hep-th]. 

\bibitem{CDMR} P.K. Concha, R. Durka, N. Merino, E.K. Rodríguez, \textit{New
family of Maxwell like algebras}, Phys. Lett. B \textbf{759} (2016) 507.
arXiv:1601.06443 [hep-th].

\bibitem{Durka} R. Durka, \textit{Resonant algebras and gravity}, J. Phys. A%
\textbf{50} (2017) 145202. arXiv:1605.00059 [hep-th].

\bibitem{CDIMR} P.K. Concha, R. Durka, C. Inostroza, N. Merino, E.K. Rodrí%
guez, \textit{Pure Lovelock gravity and Chern-Simons theory}, Phys. Rev. D
\textbf{94} (2016) 024055. arXiv:1603.09424 [hep-th],

\bibitem{CMR} P.K. Concha, N. Merino, E.K. Rodríguez, \textit{Lovelock
gravity from Born-Infeld gravity theory}, Phys. Lett. B \textbf{765} (2017)
395. arXiv:1606.07083 [hep-th].

\bibitem{CR3} P. Concha, E. Rodríguez, \textit{Generalized Pure Lovelock
Gravity}, Phys. Lett. B \textbf{774} (2017) 616. arXiv:1708.08827 [hep-th].

\bibitem{BGKL2} S. Bonanos, J. Gomis, K. Kamimura, J. Lukierski, \textit{%
Deformations of Maxwell Superalgebras and Their Applications}, J. Math.
Phys. \textbf{51} (2010) 102301. arXiv:1005.3714 [hep-th].

\bibitem{SS2} D.V. Soroka, V.A. Soroka, \textit{Semi-simple o(N)-extended
super-Poincaré algebra}, arXiv:1004.3194 [hep-th].

\bibitem{FISV} O. Fierro, F. Izaurieta, P. Salgado, O. Valdivia, \textit{%
Minimal AdS-Lorentz supergravity in three-dimensions}, Phys. Lett. B \textbf{788} (2019) 198. arXiv:1401.3697 [hep-th].

\bibitem{Bonanos:2008kr} S. Bonanos, J. Gomis, \textit{A Note on the Chevalley-Eilenberg Cohomology for the Galilei and Poincare Algebras}, J. Phys. A \textbf{42} (2009) 145206. arXiv:0808.2243 [hep-th].

\bibitem{Bonanos:2008ez} S. Bonanos, J. Gomis, \textit{Infinite Sequence of Poincare Group Extensions: Structure and Dynamics}, J.  Phys.  A \textbf{43} (2010) 015201. arXiv:0812.4140 [hep-th].

\bibitem{Gomis:2009vm} J. Gomis, K. Kamimura, J. Lukierski, \textit{Deformed Maxwell Algebras and their Realizations}, AIP Conf. Proc.  \textbf{1196} (2009) 124. arXiv:0910.0326 [hep-th].

\bibitem{Gibbons:2009me} G.W. Gibbons, J. Gomis, C.N. Pope, \textit{Deforming the Maxwell-Sim Algebra}, Phys. Rev. D \textbf{82} (2010) 065002. arXiv:0910.3220 [hep-th].

\bibitem{CMRSV} P. Concha, N. Merino, E. Rodr\'{i}guez, P. Salgado-Rebolledo, O. Valdivia, \textit{Semi-simple enlargement of the $\mathfrak{bms}_3$ algebra from a $\mathfrak{so}(2,2)\oplus\mathfrak{so}(2,1)$ Chern-Simons theory}, arXiv:1810.12256 [hep-th].


\bibitem{DAuria:1982uck} R. D'Auria, P. Fr\'{e}, \textit{Geometric Supergravity in d = 11 and Its Hidden Supergroup}, Nucl. Phys. B \textbf{201} (1982) 101.

\bibitem{Green:1989nn} M.B. Green, \textit{Supertranslations, Superstrings and {Chern-Simons} Forms}, Phys. Lett. B \textbf{223} (1989) 157.

\bibitem{Andrianopoli:2016osu} L. Andrianopoli, R. D'Auria, L. Ravera, \textit{Hidden Gauge Structure of Supersymmetric Free Differential Algebras}, JHEP {\bf 1608} (2016) 095. arXiv:1606.07328 [hep-th].

\bibitem{Andrianopoli:2017itj} L. Andrianopoli, R. D'Auria, L. Ravera, \textit{More on the Hidden Symmetries of 11D Supergravity}, Phys. Lett. B \textbf{772} (2017) 578. arXiv:1705.06251 [hep-th].


\bibitem{MO} O. Miskovic, R. Olea, \textit{Topological regularization and self-duality in four-dimensional anti-de Sitter gravity}, Phys. Rev. D\textbf{79} (2009) 124020. arXiv:0902.2082 [hep-th].

\bibitem{MOT} O. Miskovic, R. Olea, M. Tsoukalas, \textit{Renormalized AdS action and Critical Gravity}, JHEP \textbf{08} (2014) 108. arXiv:1404.5993 [hep-th].

\bibitem{ABCDFFM} L. Andrianopoli, M. Bertolini, A. Ceresole, R. D'Auria, S. Ferrara, P. Fré, T. Magri, \textit{N=2 Supergravity and N=2 super Yang-Mills theory on general scalar manifolds: Symplectic covariance, gaugings and the momentum map}, J. Geom. Phys. \textbf{23} (1997) 111. [hep-th/9605032].

\bibitem{ACDRT} L. Andrianopoli, P. Concha, R. D'Auria, E. Rodríguez, M. Trigiante, \textit{Observations on BI from $\mathcal{N}=2$ Supergravity and the General Ward Identity}, JHEP \textbf{11} (2015) 061. arXiv:1508.01474 [hep-th].

\bibitem{Ciambelli:2018wre} L. Ciambelli, C. Marteau, A.C. Petkou, P.M. Petropoulos, K.Siampos, \textit{Flat holography and Carrollian fluids}, JHEP \textbf{1807} (2018) 165. arXiv:1802.06809 [hep-th].

\end{thebibliography}
\end{document}